\documentclass[sigconf]{acmart}
\AtBeginDocument{%
  }

\setcopyright{acmlicensed}
\copyrightyear{2018}
\acmYear{2018}
\acmDOI{XXXXXXX.XXXXXXX}
\acmConference[Conference acronym 'XX]{Make sure to enter the correct
  conference title from your rights confirmation email}{June 03--05,
  2018}{Woodstock, NY}
\acmISBN{978-1-4503-XXXX-X/2018/06}

\begin{document}

\title{Integrating Large Language Models into Text Animation: An Intelligent Editing System with Inline and Chat Interaction}


\author{Bao Zhang}
\orcid{0009-0003-4161-341X}
\authornote{Equal contribution}
\email{12332459@mail.sustech.edu.cn}
\affiliation{%
  \institution{Southern University of Science and Technology}
  \city{Shenzhen}
  \state{Guangdong}
  \country{China}
}

\author{Zihan Li}
\authornotemark[1]
\email{12213030@mail.sustech.edu.cn}
\affiliation{%
  \institution{Southern University of Science and Technology}
  \city{Shenzhen}
  \state{Guangdong}
  \country{China}
}

\author{Zhenglei Liu}
\email{12112903@mail.sustech.edu.cn}
\affiliation{%
  \institution{Southern University of Science and Technology}
  \city{Shenzhen}
  \state{Guangdong}
  \country{China}
}

\author{Huanchen Wang}
\email{wanghc2022@mail.sustech.edu.cn}
\orcid{0000-0001-9339-1941}
\affiliation{%
  \institution{Southern University of Science and Technology}
  \city{Shenzhen}
  \state{Guangdong}
  \country{China}
}
\affiliation{%
  \institution{City University Of Hong Kong}
  \state{ Hong Kong}
  \country{China}
}

\author{Yuxin Ma}
\orcid{0000-0003-0484-668X}
\authornote{Corresponding author}
\email{mayx@sustech.edu.cn}
\affiliation{%
  \institution{Southern University of Science and Technology}
  \city{Shenzhen}
  \state{Guangdong}
  \country{China}
}

\renewcommand{\shortauthors}{Zhang et al.}
\renewcommand{\shorttitle}{Integrating Large Language Models into Text Animation}
\begin{abstract}

Text animation, a foundational element in video creation, enables efficient and cost-effective communication, thriving in advertisements, journalism, and social media. However, traditional animation workflows present significant usability barriers for non-professionals, with intricate operational procedures severely hindering creative productivity. To address this, we propose a Large Language Model (LLM)-aided text animation editing system that enables real-time intent tracking and flexible editing. 
The system introduces an agent-based dual-stream pipeline that integrates context-aware inline suggestions and conversational guidance as well as employs a semantic-animation mapping to facilitate LLM-driven creative intent translation.
Besides, the system supports synchronized text-animation previews and parametric adjustments via unified controls to improve editing workflow. A user study evaluates the system, highlighting its ability to help non-professional users complete animation workflows while validating the pipeline. The findings encourage further exploration of integrating LLMs into a comprehensive video creation workflow.
\end{abstract}


\begin{teaserfigure}
  \includegraphics[width=\textwidth]{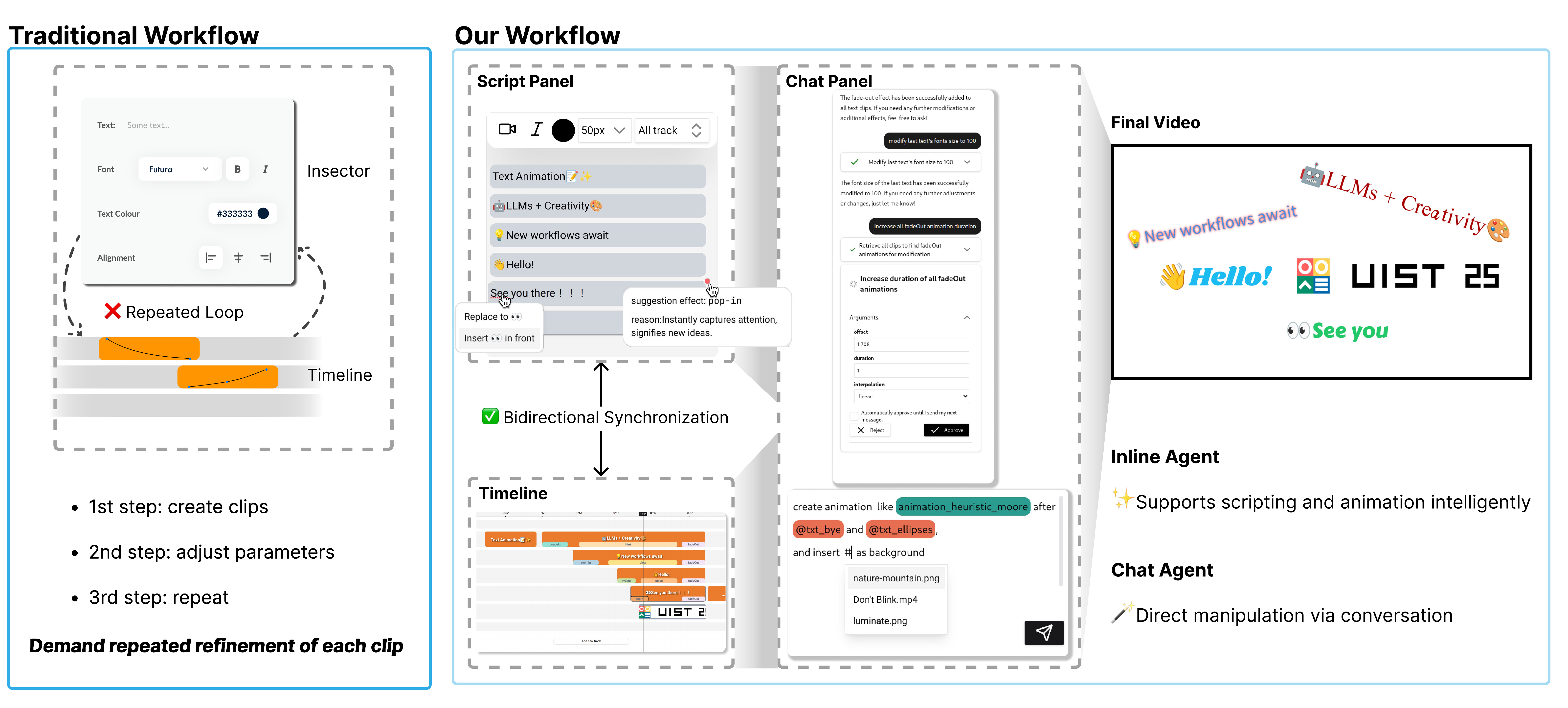}
  \caption{
    Empowering text animation creation with LLMs: Our intelligent system enables users to create and refine animations through seamless inline edits and natural chat-based interactions, blending human creativity with AI-assisted guidance.
  }
  \Description{Our system interface shows a script editor where inline agents suggest text edits and animations. Users interact through a chat panel to edit objects and refine results on a timeline and canvas.}
  \label{fig:teaser}
\end{teaserfigure}


\maketitle

\section{Introduction}
Text animation, as a key visual medium in videos, is widely utilized in contexts such as journalism and social media. 
With the proliferation of video-sharing platforms, an increasing number of creators are adopting this format to share their daily lives or produce advertisements for products. 
However, creating such videos requires specialized expertise and involves a repetitive editing workflow, making it particularly challenging for non-professional creators. Configuring text animations often demands navigating a vast library of element options, such as the animation effects, and repeatedly adjusting low-level parameters, which is both time-consuming and technically complex~\cite{Lee:2002:kinetic, Xie:2024:wordcloud}.
While rule-based methods for automatic text animation generation partially address these challenges, they rely on predefined templates and fail to capture users' diverse and nuanced intentions, thereby restricting personalization~\cite{Yeo:2008:emotional, Lee:2007:emotive, Xie:2023:wakey}.

In recent years, natural language-based approaches have been introduced into video editing domains. 
By leveraging natural language interaction, users can directly express their creative intentions without manual operations, providing a more intuitive video creation experience. Some commercially deployed projects~\cite{RunwayFrames, StabilityAI} utilize diffusion models~\cite{10.1109/TPAMI.2023.3261988, 10.1145/3626235, 10.1145/3696415,10.48550/arXiv.2408.13858,10.48550/arXiv.2407.07111} to achieve end-to-end video generation from user-provided text inputs
Beyond end-to-end generation, natural language also plays a crucial role in assisting users during the video editing process. It is employed to summarize video content and construct detailed video scripts, streamlining traditionally labor-intensive editing workflows~\cite{10.1145/3306346.3323028,10.1145/3290605.3300311,10.1145/3544548.3581494,10.1145/3379337.3415864,10.48550/arXiv.2502.10638,10.48550/arXiv.2410.03224}. Additionally, Natural Language Video Localization (NLVL) techniques enable users to extract specific video segments based on language queries~\cite{10.1145/3306346.3323028,10.1145/3290605.3300311,10.1145/3544548.3581494,10.1145/3379337.3415864,10.48550/arXiv.2502.10638,10.48550/arXiv.2410.03224}, providing a precise and efficient means to manipulate and refine video content, further enhancing the creative process.

Advancing further, Large Language Models (LLMs) have significantly improved natural language interaction in video editing workflows by demonstrating a strong capacity to understand user intent and generate contextualized responses. These capabilities enable LLMs to offer diverse content options, interpret complex queries, and provide creative suggestions for video editing. Recent studies have explored the integration of LLMs into video editing processes, facilitating tasks such as creative ideation, content and script augmentation, and clip trimming~\cite{10.1109/GEN4DS63889.2024.00008, 10.1109/GEN4DS63889.2024.00008, 10.1145/3640543.3645143, 10.1145/3643834.3661591, 10.1145/3708359.3712090}.
Despite the advancements brought by LLMs, current LLM-aided video editing systems face two major challenges. First, systems that rely on LLMs for direct content generation often produce uncontrollable outputs due to hallucinated content, limited domain knowledge, and restricted user control. Second, agent-based systems using LLMs for guidance or recommendations lack support for editing fine-grained elements, such as font attributes or animation effects. These challenges highlight the need to integrate LLM capabilities with tools that allow for precise, user-driven editing, improving both flexibility and reliability in video editing workflows.

To address these challenges, we propose an LLM-aided text animation editing system designed to lower creative barriers for non-professional users and enhance editing efficiency. For improving the LLM's understanding of user intent and fostering better editing, the system employs a semantic-animation mapping to translate user inputs into meaningful video editing actions. It also incorporates a dual-mode agent-based pipeline: the \textbf{Inline Agent}, which dynamically recommends animation effects based on textual semantics, and the standalone \textbf{Chat Agent}, which facilitates natural language interaction and drag-and-drop referencing, with user validation ensuring accurate batch editing. Both of them enhance the reliability and granularity of LLM assistance and the flexibility of editing. Regarding the traditional editing workflow, the system simplifies processes through a bidirectional linkage mechanism between the script and timeline panels.
The script panel adopts a rich-text-like design for line-by-line text attribute adjustments, while the timeline panel features video-editing-style track management. Intelligent mapping strategies synchronize these panels, reducing redundant operations and optimizing the editing process.

To evaluate our system, we conducted a user study with 11 participants, which included diverse backgrounds video editors, to assess the effectiveness in aiding text animation editing.
The results demonstrated that participants were able to produce satisfactory AI-collaborative text animation outcomes using our system.
Users expressed positive feedback regarding the system's functionalities, finding them easy to use and helpful for producing creative text-based video artifacts
They also appreciated our work on agent integration and the two modes of interaction with agents.
Additionally, our study revealed the influence of LLMs on existing workflows and illustrated how different types of agents can support distinct tasks within the workflow.
Our system can serve as a reference for integrating LLMs into future multimedia content creation tools and inspire subsequent research.

In summary, this paper makes the following contributions:
\begin{itemize}
    \item A \textbf{formative study} involving nine participants that identifies the workflow and challenges in text animation editing.
    \item A \textbf{interactive system} for text animation editing, which supports more flexible and granular editing through LLM-power agents and enhanced workflow.
    \item The \textbf{design of semantic-animation mapping and agent-based dual-mode pipeline} ensures the capability of our system to understand editing intent and element suggestion.
    \item A \textbf{user study} validating the pipeline and workflow design, while demonstrating the usefulness of the text animation editing system.
\end{itemize}

\section{Related Work}
\subsection{Human-AI Collaboration}

Since the public release of GPT in 2022, Large Language Models (LLMs) have played a pivotal role in human productivity and daily life. Since then, researching how to optimize LLMs for enhanced human service has remained a critical academic focus. Recent studies suggest that expecting current LLMs to solve problems fully autonomously is suboptimal. Instead, a more effective approach involves designing interactive applications where AI Agents collaborate with humans to accomplish tasks~\cite{10.1016/j.chb.2022.107502, 10.1016/j.jbusres.2020.11.038, 10.1145/3173574.3174223, 10.1007/978-3-030-78462-1_13, 10.17705/1jais.00867}.

When designing AI-human interaction systems, essential design principles must be observed to improve user experience and system performance. First, AI Agents require enhanced perception capabilities~\cite{10.1145/3613904.3642129} and configurable user options \cite{10.1145/3613904.3642868} to differentiate between users~\cite{10.1016/j.ijhcs.2024.103301} and deliver personalized outputs that address diverse needs. Second, AI systems should clearly communicate their capabilities and limitations to users~\cite{10.1007/s10796-022-10284-3, 10.1145/3579612}, enabling rational task allocation between human and AI counterparts. This mitigates issues of over-reliance or underutilization of AI capabilities while optimizing collaborative efficiency. Finally, such systems must maintain human oversight by transparently revealing the AI's reasoning process during output generation~\cite{10.1007/s10796-022-10284-3, 10.1016/j.ijhcs.2024.103301}. Critical operational steps should allow human intervention, fostering user trust in AI-integrated systems and enabling quality improvement through human participation when AI approaches its capability boundaries.

In this study, we focus on the "text-to-video editing" task by integrating LLMs into the workflow. Our LLM-powered agent not only monitors the editor's global state and user-specified elements, but also interprets manual operations to infer editorial intent, subsequently providing personalized editing recommendations. During suggestion generation, it transparently presents its reasoning process. The system additionally enables users to query the agent's capability list, facilitating more effective human-AI collaboration.

\subsection{Video Editing}

Video, with its multi-sensory integration of visual and auditory elements, has consistently served as the most efficient mainstream medium for information dissemination. Professional editing software such as Premiere Pro~\cite{AdobePremierePro} and DaVinci Resolve~\cite{DaVinciResolve}, despite their comprehensive functionality in film production, present steep learning curves that demand substantial time investment from casual users. While amateur-oriented tools like CapCut~\cite{CapCut} significantly reduce learning barriers, they remain functionally incomplete for specialized video editing tasks.

Recent advancements in AI technology have spurred growing research into AI-assisted video creation. Current AI editing tools enable basic operations like trimming and sequencing through text instructions via LLMs. For instance, Lave~\cite{10.1145/3640543.3645143} leverages LLMs to generate conceptual suggestions, storyboard planning, and automated editing, yet remains constrained by its reliance on existing video clips and tight audio-visual coupling. Lotus~\cite{10.1145/3708359.3712090} addresses this flexibility limitation by extracting highlights from long-form videos and enabling text-based narration modifications. However, its visual modifications are restricted to cropping and rearrangement without substantive content alteration. PodReels~\cite{10.1145/3643834.3661591} specializes in identifying highlights from lengthy podcast videos and compiling them into short clips with transitional effects and captions, though primarily retaining original visual content.

On the other hand, several studies have explored structured data-driven video generation. Amini et al.~\cite{10.1145/2702123.2702431} analyzed data-driven videos and summarized a narrative structure comprising ``establisher, initial, peak, release.'' WonderFlow~\cite{10.1109/TVCG.2024.3411575} investigated narrative-driven approaches for data video generation, while DataPlayer~\cite{10.1109/TVCG.2023.3327197} and Data Director~\cite{10.1109/GEN4DS63889.2024.00008} achieved end-to-end generation of data-driven videos by integrating the semantic comprehension capabilities of LLMs. However, current efforts inadequately implement Human-AI collaborative generation of data videos. The generation quality remains constrained by either the user's professional expertise or the LLM's interpretive capabilities, failing to fully leverage the combined potential of human creativity and AI efficiency.


Building upon these foundations, we propose a dedicated LLM-aided text animation editing system. Our LLM-powered agent supports parametric adjustments for each text animation element, enabling precise control over visual outcomes beyond mere reliance on source material effects. Notably, contemporary video editing research predominantly focuses on diffusion model-based text-to-video generation~\cite{10.48550/arXiv.2303.04761, 10.48550/arXiv.2303.12688}. While these innovations provide novel material creation methods for professionals, our work emphasizes optimizing the editing process for existing visual assets in text-centric video production.

\subsection{Direct Manipulation}
The Direct Manipulation framework proposed in the 1980s~\cite{10.1080/01449298208914450} significantly outperformed traditional command-based interaction in intuitiveness, real-time feedback, and low learning threshold, driving the development of early GUI systems. This interaction framework greatly enhanced the intuitiveness of human-computer interaction systems while boosting user confidence and satisfaction~\cite{10.1145/1015579.810991, 10.1145/62402.62428}. Early research also proposed multimodal interface designs combining natural language with direct manipulation techniques~\cite{10.1145/67450.67494}, introducing simple methods for natural language-based interface control through graphical context management, focus buttons, and logical form constraints, thereby improving user efficiency and reducing cognitive load. However, these early studies were limited by the NLP capabilities of the time in handling complex instructions and decision-making processes, resulting in relatively simplistic functionality.

Recent research has integrated the Direct Manipulation framework with cutting-edge AI achievements to achieve better human-computer interaction experiences in more complex tasks. For instance, SwapVid~\cite{10.1145/3613904.3642515} combines this framework with OCR technology, establishing bidirectional synchronization between video timelines and document positions by analyzing text content in video frames and documents, significantly reducing users' physical and mental demands while improving system usability. DirectGPT~\cite{10.1145/3613904.3642462} integrates Direct Manipulation with LLMs, maintaining real-time visualization and physical interactions (e.g., drag-and-drop) with LLM-generated objects, substantially enhancing operational efficiency and user experience in specific tasks.

In our system, we similarly employ the Direct Manipulation framework to build our visual interaction experience. As our system is based on timeline-oriented traditional video editing tools like CapCut, it inherently maintains continuous display of generated objects. Meanwhile, we designed an LLM interaction pattern that combines graphical context management with rich physical interactions. This design clearly manages user-AI Agent interaction history while allowing users to freely select page elements to send inline in Agent inputs through drag-and-drop or @mention operations. This enables users to intuitively invoke AI Agents, reduces learning costs and mental burden, and enhances system usability.

\section{Formative Study}
To inform our system design, we conducted a formative study with 9 amateur or semi-professional video creators. The study aimed to:

\begin{itemize}
\item Understand current video creation workflows to identify the needs of non-professional video creators for editing tools;
\item Gather their feedback and experiences regarding AI-assisted video editing to guide the functional and interaction design of the system’s AI Agent.
\end{itemize}

\subsection{Methodology}
\subsubsection{Participants}
We recruited nine participants (6 male, 3 female) from universities and enterprises. All participants were amateur or semi-professional creators with video creation experience, averaging 2.3 years of video production experience. The study was conducted through semi-structured interviews lasting approximately one hour, with all participants receiving 100 RMB (approximately 14 USD) remuneration upon completion. The basic information of participants is as follows:

\begin{table}[h!]
\centering
\caption{Participant information in formative study}
\begin{tabular}{|c|c|c|c|c|}
\hline
\textbf{Participant} & \textbf{Age} & \textbf{Gender} & \textbf{Identity} & \textbf{Years of Experience} \\ \hline
P1 & 24 & Male & Student & 4 \\ \hline
P2 & 23 & Female & Staff & 1 \\ \hline
P3 & 20 & Male & Student & 2 \\ \hline
P4 & 19 & Male & Student & 3 \\ \hline
P5 & 18 & Male & Student & 3 \\ \hline
P6 & 23 & Female & Staff & 2 \\ \hline
P7 & 25 & Female & Staff & 3 \\ \hline
P8 & 19 & Male & Student & 2 \\ \hline
P9 & 20 & Male & Student & 1 \\ \hline
\end{tabular}
\label{tab:participant_info}
\end{table}

\subsubsection{Procedure}
With participants' consent, we first introduced the background of text animation editing, clarified our research objectives, and explained the intent of the formative study.

First, we collected background information about the participants, including gender, age, social identity, years of experience, commonly used video editing tools, and other relevant details.
Second, we asked the participants to demonstrate their typical workflow when using their preferred video editing tools for content creation and to elaborate on the pain points they encountered during usage.
Third, participants were asked to watch four representative text animation videos (each surpassing 500,000 views) and engage in a discussion about their expectations for the attributes of text animations, as well as their views on the functionality and interaction methods of LLM-assisted text animation editing tools.
Finally, we engaged in an open-ended discussion with the participants to document their personal perspectives on specific issues.

\subsection{Results}
\subsubsection{Current Video Editing Workflows}
During the creative process, users first prepare the required video materials and the script/text content for the video, then construct the initial version of the video. After multiple iterations, they achieve the desired outcome and complete the video creation. Text-related elements such as subtitles may be added either during the initial video construction phase or after finalizing video modifications, depending on user preferences. Regarding material management, users primarily organize assets manually through computer folders. In terms of template usage, the majority of our interviewees tend not to employ templates for creation. Specific details are illustrated in the following diagram:

\begin{figure}[!htbp]
\centering
\includegraphics[width=0.5\textwidth]{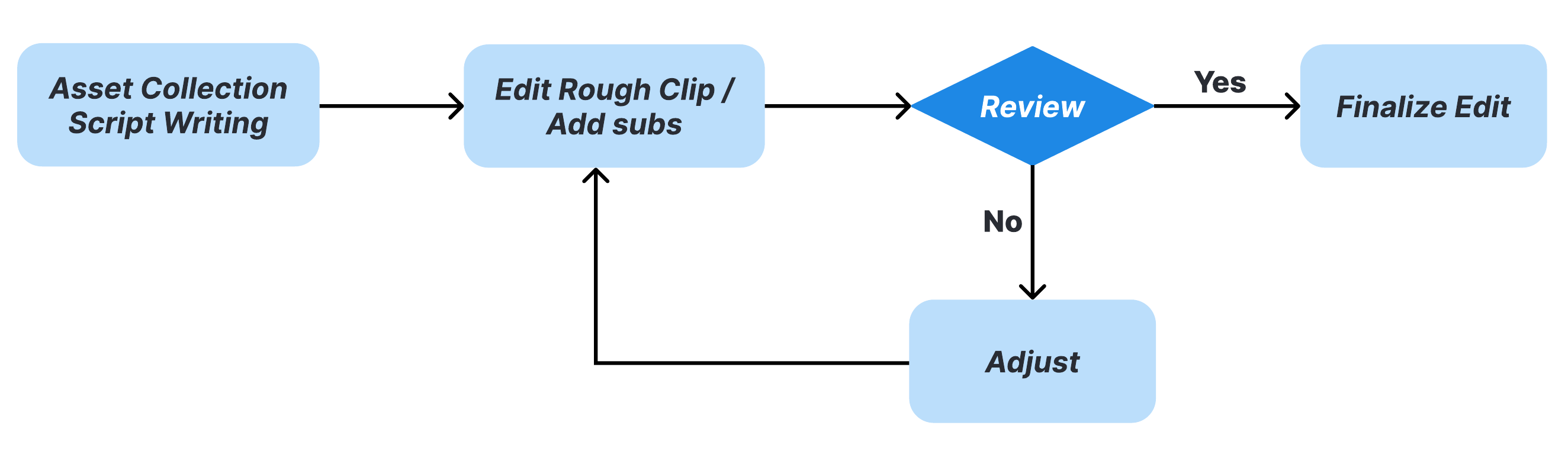}
\caption{Illustration of Current Video Editing Workflows}
\label{fig:figure/traditional_workflow}
\vspace{-.46cm}
\end{figure}

\subsubsection{Drawbacks in Current Video Creation Workflows}
\begin{itemize}
    \item Balance between Professionalism and Usability: Professional video editing software like Adobe Premiere, while offering robust editing capabilities, suffers from complex workflows and steep learning curves that make them unsuitable for casual users. Conversely, simplified tools like CapCut streamline the editing process and reduce learning costs but fall short in addressing advanced editing requirements due to their limited feature sets.
    \item Inability to Understand Editing Intent for Elements Suggestion: While contemporary video editing tools, including those powered by LLMs, offer auxiliary features such as material recommendations to assist users in constructing visual elements, user feedback highlights significant limitations in their effectiveness. These tools often struggle to accurately interpret user intent and frequently produce suboptimal output quality. As a result, they fail to meet user demands for high-quality visual elements, leaving users dependent on manually sourcing and curating materials for their video content.
    \item Cumbersome Text Editing Processes: Existing video editors require tedious manual operations for inserting subtitles and adjusting timelines, demanding manually inserting each individual subtitle block and repeated refinements to achieve satisfactory results. Automatic subtitle generation tools are available but often lack accuracy, requiring manual verification and adjustment to ensure reliability. Furthermore, though some tools employ end-to-end models allowing natural language-based content modifications, it lacks support for fine-grained adjustments to specific output segments. This limitation restricts user control and proves inadequate for detail-oriented editing tasks.
\end{itemize}

\section{Design Goal and Consideration}

Based on the formative study, we proposed design goals to guide the interaction design for text animation creation. We further made a series of design considerations to realize these goals. 
In this section, we introduce both the design goals and design considerations.

\subsection{Design Goals}
Our formative study revealed several issues within current video editing workflows. Based on these findings and gaps identified in related work, we established design goals to guide the design and development of an LLM-aided system for text animation editing. This aims to achieve high-level human-AI co-editing capabilities while providing amateur and semi-professional users with advanced features and low learning costs.
\begin{itemize}
    \item[\textbf{DG 1}] \textbf{Improving Usability While Ensuring Functionality:} This goal integrates agent-based assistance and multi-granularity editing to balance advanced functionality with ease of use. LLM-powered agents proactively infer user intent from manual edits, recommending operations and providing automated project scaffolding to reduce complexity and inspire novice users. Simultaneously, multi-granularity editing ensures user involvement, control, and flexibility, supporting both adjustments and refinements.
    \item[\textbf{DG 2}]\textbf{Enhancing System Understanding of Editing Intent:} This goal aims to improve the system's ability to understand user intent by combining an agent-based dual-mode pipeline with Text-to-Animation Semantic Mapping. The dual-mode pipeline facilitates better interaction between the agent and the user, allowing the system to perceive context and offer guidance throughout the editing process. This ensures a more accurate understanding and translation of user intent. Additionally, the semantic-animation mapping links linguistic features, such as semantic importance and emotional tone, with animation parameters to generate suggestions that align more closely with user expectations and creative goals.
    \item[\textbf{DG 3}] \textbf{Creating an Intuitive Text Animation Editing Process:} This goal focuses on making text animation editing more intuitive by redesigning traditional workflows to integrate the script editor directly into the video editor. A dedicated text animation workspace was developed to co-locate textual content alongside its corresponding animations. Each animation is visually represented by an icon that provides a direct preview of its effect, enabling users to intuitively modify text-animation mappings. 
\end{itemize}

\subsection{Design Considerations}

\subsubsection{Integrating LLMs and Improving the LLM-Aided Mechanism in Video Editing \textbf{(DG1, DG2, and DG3)}}
In the formative study, we observed that the majority of participants were already utilizing LLMs, such as ChatGPT, to optimize their scripts. All participants expressed a strong desire for LLM integration within the text animation creation workflow. They highlighted the potential of LLMs not only to better understand creative intent but also to provide suggestions for animation elements and editing decisions. Such integration would reduce the manual effort required for novice users while also enabling a multi-level, LLM-assisted workflow that supports flexible and adaptable editing processes for users with varying levels of expertise.

Furthermore, to ensure the capability of LLM in understanding editing intent and element suggestion, we first summarized the semantic-animation mapping space based on the attributes mentioned by participants and the literature review~\cite{10.1109/TVCG.2024.3411575, Xie:2024:wordcloud, Shao:2025:animation}, which connects linguistic features (semantic importance, emotional tone) with animation parameters. This includes:
1) \textbf{Static attributes}: Font size, color, positioning;
2) \textbf{Dynamic behaviors}: Animation type, velocity, temporal patterns. This structured representation improves LLMs' comprehension of textual-visual relationships for more context-aware suggestions.
Based on the mapping space, we introduce dual-mode LLM-power agents for either context-specific interactions with the user for suggestions or provide conversation interaction with the user as guidance, which preserves the editing functionalities while enhancing the usability of the system through LLM-based assistance.
    
\subsubsection{Preset Animation vs. Keyframe Animation \textbf{(DG1 and DG3)}}
Based on our formative study, we focused on two primary approaches for animation: \textbf{Preset Animation} and \textbf{Keyframe Animation}. 

\textbf{Preset Animation:} This approach utilizes predefined animation templates that allow users to quickly apply animations to content. 
Each template provides a specific style or effect, such as movement, scaling, or fading. 
Users can adjust high-level parameters (e.g., speed, duration, direction) to fit their creation needs. 
The main advantage of preset animation is its simplicity—users don’t need to manually define each animation keyframe, making it accessible for non-professionals. 
This method is especially useful for quick, visually appealing results without a steep learning curve.

\textbf{Keyframe Animation:} In contrast, keyframe animation provides greater flexibility, allowing users to define specific points in the animation timeline (keyframes) and adjust clip properties (e.g., position, rotation, opacity) at each keyframe. This method offers full creative control over the animation process, allowing for highly customized and intricate animations. 
However, keyframe animation is more complex and time-consuming, requiring a deeper understanding of animation principles. 
This makes it less suitable for novice users who may feel overwhelmed by the process.

Guided by the findings from our formative study, we chose to implement Preset Animation in our system. 
The study highlighted that non-professional users have a low tolerance for complexity and prefer quick, intuitive workflows.
By focusing on Preset Animation, we aim to provide a user-friendly editing experience that aligns with these needs for simplicity and efficiency.
Additionally, through mapping space, a connection can be established between user intent and recommended animations.

\begin{figure*}[!htbp]
\centering
\includegraphics[width=0.9\textwidth]{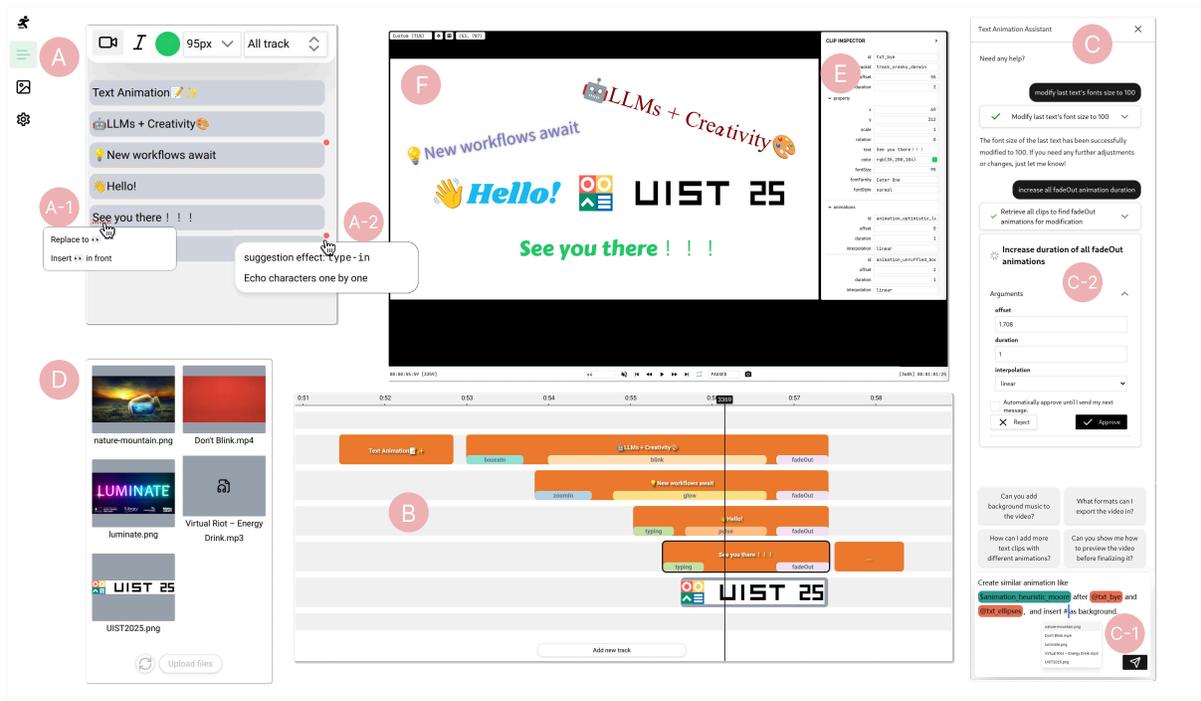}
\caption{
User Interface Overview: The interface consists of six panels: Script Panel (A), Timeline Panel (B), Chat Panel (C), Resource Panel (D), Inspector Panel (E), and Preview Panel (F). 
These components work together to provide a comprehensive and intuitive video editing system, enabling users to seamlessly interact with content, manage resources, and preview edits in real time.
}
\vspace{-.26cm}
\label{fig:system_overview}
\end{figure*}

\subsubsection{Bridging the Gap between Script Creation and Video Production \textbf{(DG3)}}
In traditional workflows, the process of writing a script and editing a video is often separate, leading to fragmented and inefficient creative experience.
This separation creates a gap between the conceptualization of content (scriptwriting) and its actual visualization (video editing), making it harder for creators to maintain consistency and synchronize narrative with visual elements. While the separated workflow works well in professional video production environments—where multiple people with specialized roles can benefit from a shared platform—we recognize that it may not always be the ideal solution for amateur or independent creators. 

Based on the findings from user interviews, we discovered that for such creators, particularly when frequent adjustments to both the script and video content are required, the traditional separation between script and video editing can negatively impact the user experience.
Therefore, we decided to integrate the script editor directly into the video editor. 
This unified approach streamlines the creative process, enabling users to see the real-time impact of script changes on the animation and video elements. 
We expect this approach to offer a more cohesive and intuitive experience, where both textual content and visual elements are edited in tandem, ensuring better alignment between the narrative and its visual representation.

\section{User Interface}
\begin{figure}[!htbp]
\centering
\includegraphics[width=0.49\textwidth]{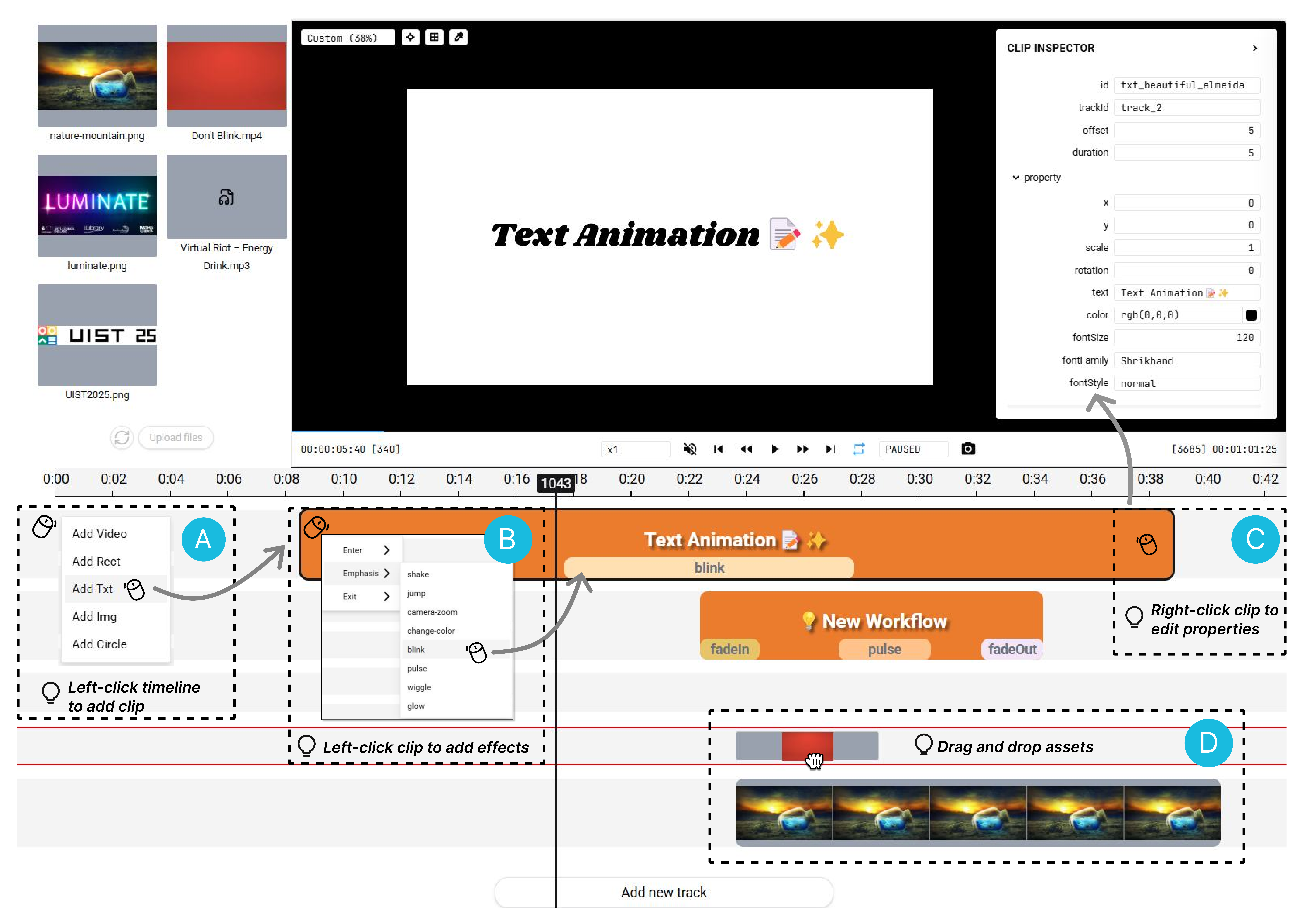}
\caption{Basic Editing Capabilities: Users can create clips by right-clicking on a track in the timeline (A) or by directly dragging resources onto the timeline (D). Right-clicking on a clip opens a context menu where users can create animations (B). Additionally, users can select clips or animations by click to inspect and adjust their properties through the Inspector Panel (C), enabling precise control over the content.}
\label{fig:edit_basic}
\vspace{-.46cm}
\end{figure}

Guided by the design goals and considerations, we developed the system which comprises six primary panels: 1) Script Panel, 2) Timeline Panel, 3) Chat Panel, 4) Resource Panel, 5) Inspector Panel, and 6) Preview Panel, as shown in~\autoref{fig:system_overview}. These panels establish the foundation for the system's core editing functionalities (\textbf{DG1}), with a detailed focus on the collaborative Script-Timeline editing for a refined workflow (\textbf{DG 3}) and the conversational interface with a dual-mode pipeline to leverage LLM capabilities effectively (\textbf{DG1 and DG2}).

\subsection{Basic Editing Interface}

We have implemented basic video and animation editing capabilities (\autoref{fig:edit_basic}), creating a basic editing interface consisting of the timeline, inspector panel, preview panel, and resource panel.

\begin{figure*}[!htbp]
\centering
\includegraphics[width=0.9\textwidth]{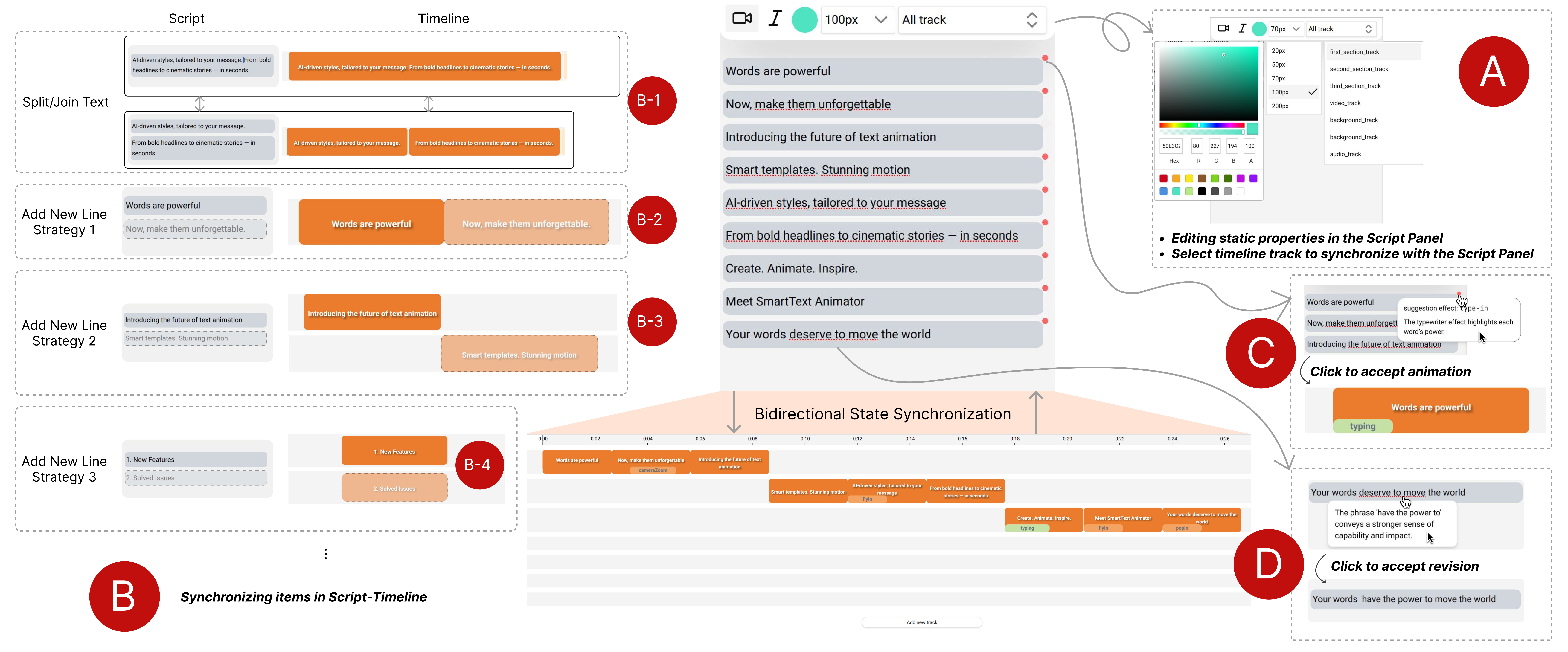}
\caption{Script-Timeline State Synchronization and Editing Capabilities. The script editor and timeline remain synchronized, with any changes in the script automatically reflected in the corresponding timeline clips.
The script editor provides real-time text revision suggestions and animation recommendations, helping users efficiently enhance their content. These suggestions can be approved or modified directly, streamlining the editing process.
}
\vspace{-.26cm}
\label{fig:script_timeline}
\end{figure*}

\subsubsection{Resource Panel (\autoref{fig:system_overview} D)}
The Resource Panel houses all available assets for the project, including images, audio, and videos. 
To upload new resources, users can click the upload button, which allows them to import additional media into the panel. 
Users can easily access and drag them onto the timeline for editing (\autoref{fig:system_overview} B).
Additionally, users can drag and drop resources into the Chat Panel for reference and further manipulation within the conversation (\autoref{fig:system_overview} C-1).

\subsubsection{Timeline Panel (\autoref{fig:system_overview} B)}
The Timeline Panel is where users can arrange and organize clips (text, video, images, audio, and elements) along a horizontal track. 
It allows users to define the order and duration of each clip, enabling precise control over the sequencing of content. 
Through the timeline, users can visually track the progression of their multimedia elements and make adjustments as needed.
Right-click menus allow users to add clips and preset animations, while the drag-and-drop functionality provides a simple way to reposition clips along the timeline.

\subsubsection{Inspector Panel (\autoref{fig:system_overview} E)}
The Inspector Panel provides users with granular control over the properties of clips and animations. 
Here, users can adjust key settings such as position, scale, opacity, rotation, and other animation-related parameters. 
This panel enables fine-tuned customization of the media elements, providing precise control over each clip's behavior and appearance, which offering flexibility and accuracy in editing.

\subsubsection{Preview Panel (\autoref{fig:system_overview} F)}
The Preview Panel allows users to visualize the edited content in real time, offering a dynamic view of how the video will look after adjustments. 
This panel supports playback, allowing users to review their work and ensure everything is in place. 
Through the Preview Panel, users can immediately see the results of any changes made in the Timeline or Inspector Panels, providing an essential feedback loop for effective editing. 
Users can also drag and adjust clips directly within the Preview Panel for further refinement.

\subsection{Script-Timeline Synchronization Editing}

\subsubsection{Script Editing Interface}
The Script Editing Interface allows users to create, edit, and refine video scripts in a familiar and efficient environment. Users can edit text as easily as they would in common document editors like Microsoft Word. In addition to basic text input and revision, users can modify static text properties—such as font style, font size, and text color—through the toolbar in the script panel (\autoref{fig:script_timeline} A). Users can select multiple lines of text to apply batch modifications, significantly improving editing efficiency when adjusting styles for different sections of the script. 

Additionally, users can choose which tracks' text clips to display within the script panel  (\autoref{fig:script_timeline} A). 
This selective viewing feature enables users to focus on specific clips or layers when editing complex projects involving multiple text, image, or element tracks, reducing visual clutter and streamlining the scripting workflow.

All changes made to the text’s appearance are immediately reflected in the Preview Panel, providing real-time feedback on the visual effects of edits. This seamless editing experience ensures that users can maintain full control over both the content and its visual presentation, setting a strong foundation for subsequent animation and synchronization steps.



\begin{figure*}[!htbp]
\centering
\includegraphics[width=0.9\textwidth]{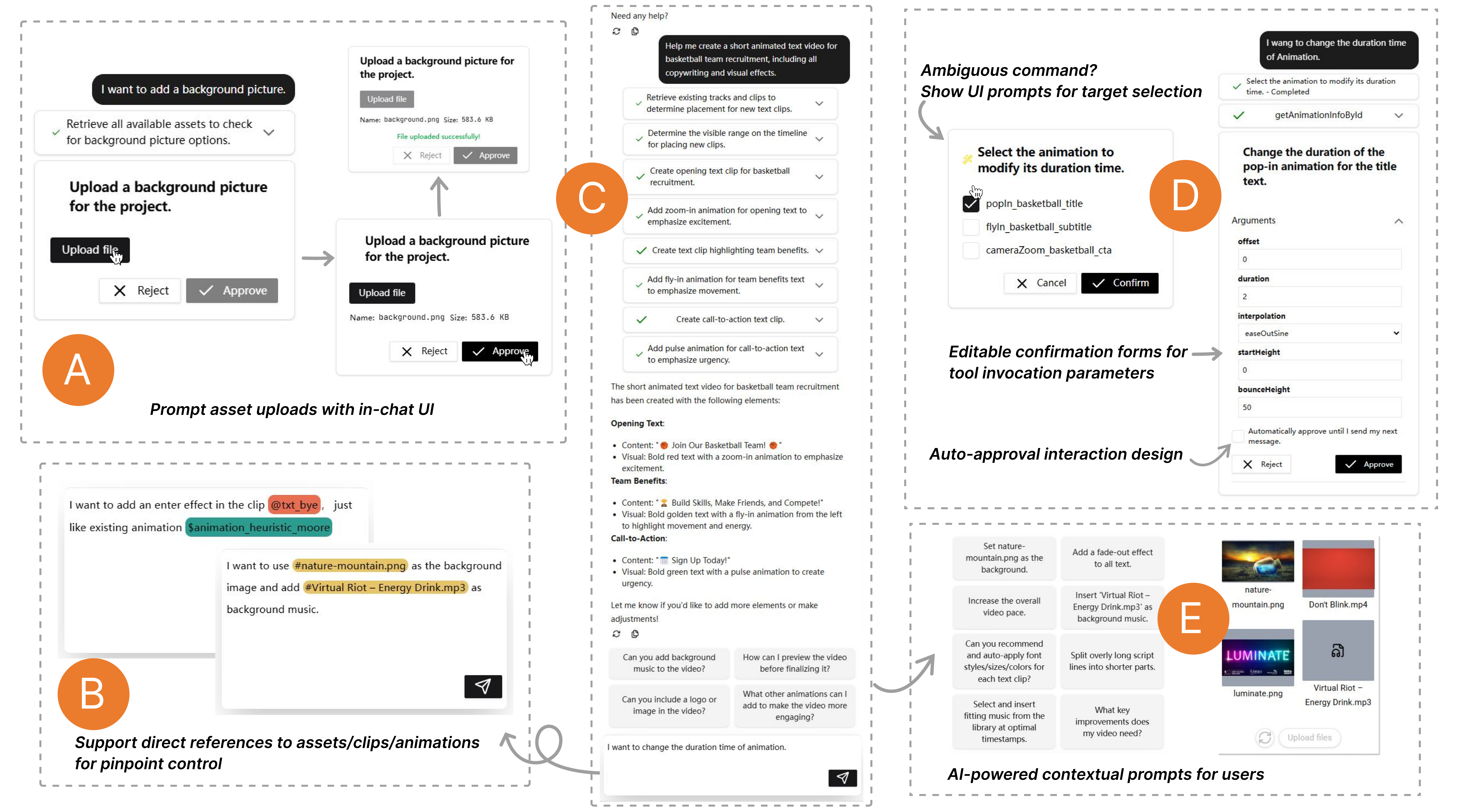}
\caption{
Chat Panel: Natural Language Video Editing and Guided Input Interaction. 
The Chat Panel enables users to edit videos directly using natural language commands. 
Users can reference and modify editable objects (clips, animations, and assets) through a direct manipulation mode. 
Additionally, guided input interaction is provided for ambiguous commands, offering context-aware suggestions based on the editor's current state to improve command clarity and execution.
}
\label{fig:chat_pane}
\vspace{-.26cm}
\end{figure*}

\subsubsection{Script-Timeline State Synchronization (\autoref{fig:script_timeline} B)}

The Script-Timeline State Synchronization ensures that changes made in the script editor are automatically reflected in the timeline, and vice versa. Any update to the script, such as text modifications, is instantly mirrored in the corresponding timeline clips, allowing users to visualize how their script changes will affect the video. 
\begin{itemize}
\item \textbf{Timeline to Script}: When updating from the timeline to the script editor, all clips in the selected track (\autoref{fig:system_overview} A) are sorted according to their appearance time. This ensures that the text content in the script editor matches the sequential order of the clips on the timeline, maintaining consistency and coherence in the editing process.
\item \textbf{Text Content Modification}: When users edit the text content in the script editor, the corresponding clips on the timeline are automatically updated to reflect the changes in real time.
\item{\textbf{Text Splitting and Joining (\autoref{fig:script_timeline} B-1)}: Users can split or merge text segments within the script editor. When splitting, the system automatically adjusts the timeline clips according to the lengths of the resulting text segments, ensuring a proportionate distribution. Merging two segments correspondingly combines their associated timeline clips.}
\item{\textbf{Add Line (\autoref{fig:script_timeline} B-2, B-3, B-4)}}: When users insert new text in the script editor, the system automatically generates a corresponding new clip on the timeline. The placement of the new clip is determined by an agent, which uses contextual information and predefined layout strategies to decide its optimal position. As shown in Figure 4(B), the predefined layout strategies include: (1) Sequential insertion on the same track (\autoref{fig:script_timeline} B-2),
(2) Parallel insertion with adjusted timing (\autoref{fig:script_timeline} B-3), (3) Parallel insertion on a new track (\autoref{fig:script_timeline} B-4), 
\end{itemize}



These state synchronization interactions are sufficient to support the majority of typical script editing workflows, effectively ensuring a smooth and consistent editing experience across panels.
However, due to the inherent differences between the interaction paradigms of the script editor and the timeline panel, our synchronization mechanism is not yet fully complete.
We will further discuss these limitations and potential improvements in subsequent sections.

\subsubsection{Script LLMs-Assist and Animation Recommendations}

In the script editor, our system provides two types of suggestions: script revision suggestions and animation recommendations. 

\begin{itemize}
    \item \textbf{Animation Recommendations (\autoref{fig:script_timeline} C):} Text that could benefit from animation will have a red dot in the top-right corner. Hovering over the red dot will reveal a menu, and by clicking on the menu, the recommended animation will be applied.
    \item \textbf{Script Revision Suggestions (\autoref{fig:script_timeline} D):} When there are text revisions suggested, the corresponding text will be underlined with a red dashed line. Upon hovering over the underlined text, a menu appears. By clicking on the menu, users can apply the suggested revision. 
\end{itemize}
All suggestions, whether for text revisions or animations, include a reason to ensure that users have a clear understanding of the proposed changes before they are applied.

\subsection{Conversational Interface}



We introduce a Conversational Interface (\autoref{fig:chat_pane}), enabling users to control and modify their content through natural language commands. This interface offers a more flexible and efficient way to interact with the system. The Conversational Interface is designed around a Plan-and-Execute task flow, where users can issue commands, and the system generates a sequence of tasks to be executed  (\autoref{fig:chat_pane} C). In the following section, we will describe how the Conversational Interface enhances user interaction.


\subsubsection{Direct Manipulation}
The Direct Manipulation feature in the Chat Panel empowers users to interact with editable objects (clips, animations, and  assets) using natural language commands. By leveraging this mode, users can reference and modify editable objects directly within the interface. 

We provide users with two methods for referencing elements to directly manipulate editable object: drag-and-drop referencing and special symbol referencing.

 \begin{itemize}
    \item \textbf{Drag-and-drop}: Based on the design philosophy of Direct Manipulation, we implemented a straightforward drag-and-drop method for referencing elements in the Chat Panel. Users can directly drag resources, timeline elements, etc., from the editor into the input text box to achieve inline element referencing. This form of referencing aligns with user interaction habits, offering simplicity and reducing learning costs.
    \item \textbf{Special symbol autocompletion (\autoref{fig:system_overview} C-1 and \autoref{fig:chat_pane} B)}: Mimicking the "@" symbol usage in chat tools, we designed a special symbol-based referencing method in the input box.  Different special symbols represent different types of editor elements. After entering a special symbol, a corresponding element list appears for user selection. This method eliminates the need for users to shift focus to other parts of the editor, enhancing system consistency.
\end{itemize}
    
   

\subsubsection{Show UI prompt for ambiguous}

In cases where a user's natural language command is ambiguous or lacks sufficient context for precise execution, the system provides a Show UI Prompt to guide the user in clarifying their intent. This feature ensures that the user can continue interacting with the system without confusion, promoting a smooth and efficient editing process.

When the system detects an ambiguous command, it dynamically presents an appropriate UI element to help the user specify their action. These UI prompts can include:

\begin{itemize}
\item \textbf{Selectors (\autoref{fig:chat_pane} D left):} Drop-down menus or selection boxes to choose from available options, such as selecting a specific clip, animation type, or transition effect.
\item \textbf{Parameters Form (\autoref{fig:chat_pane} D right):} Interactive fields for entering specific details, like adjusting text properties (font, size, color) or specifying animation parameters.
\item \textbf{Upload Buttons (\autoref{fig:chat_pane} A):} Buttons for uploading new resources or media directly into the workspace, assisting users in adding assets they may wish to use in their edits.
\end{itemize}
These UI prompts are designed to be context-sensitive, appearing based on the current task and the system's understanding of the user's needs. By incorporating these prompts, the system ensures that users can make informed decisions, providing a clearer path to executing their commands with accuracy.

\subsubsection{Context-aware instruct suggestions (\autoref{fig:chat_pane} E)}

Our system leverages advanced context-awareness to enhance user interaction through intelligent instruction suggestions. By analyzing the current editing context—such as the script content, assets timeline status, and user actions—the system offers tailored suggestions that align with the user’s workflow and goals.

\section{Backend System}
\subsection{Agent Design}
To ensure the reliability and granularity of LLM-powered agent assistance while maintaining editing flexibility, we propose a dual-mode agent-based pipeline, which includes the \textbf{Inline Agent}, which dynamically recommends animation effects based on textual semantics, and the standalone \textbf{Chat Agent}, which enables natural language interaction and drag-and-drop referencing. The pipeline, supported by user interaction and validation, ensures the rationality and controllability of the editing processes aligned with user intent.

\begin{figure}[!htbp]
\centering
\includegraphics[width=0.5\textwidth]{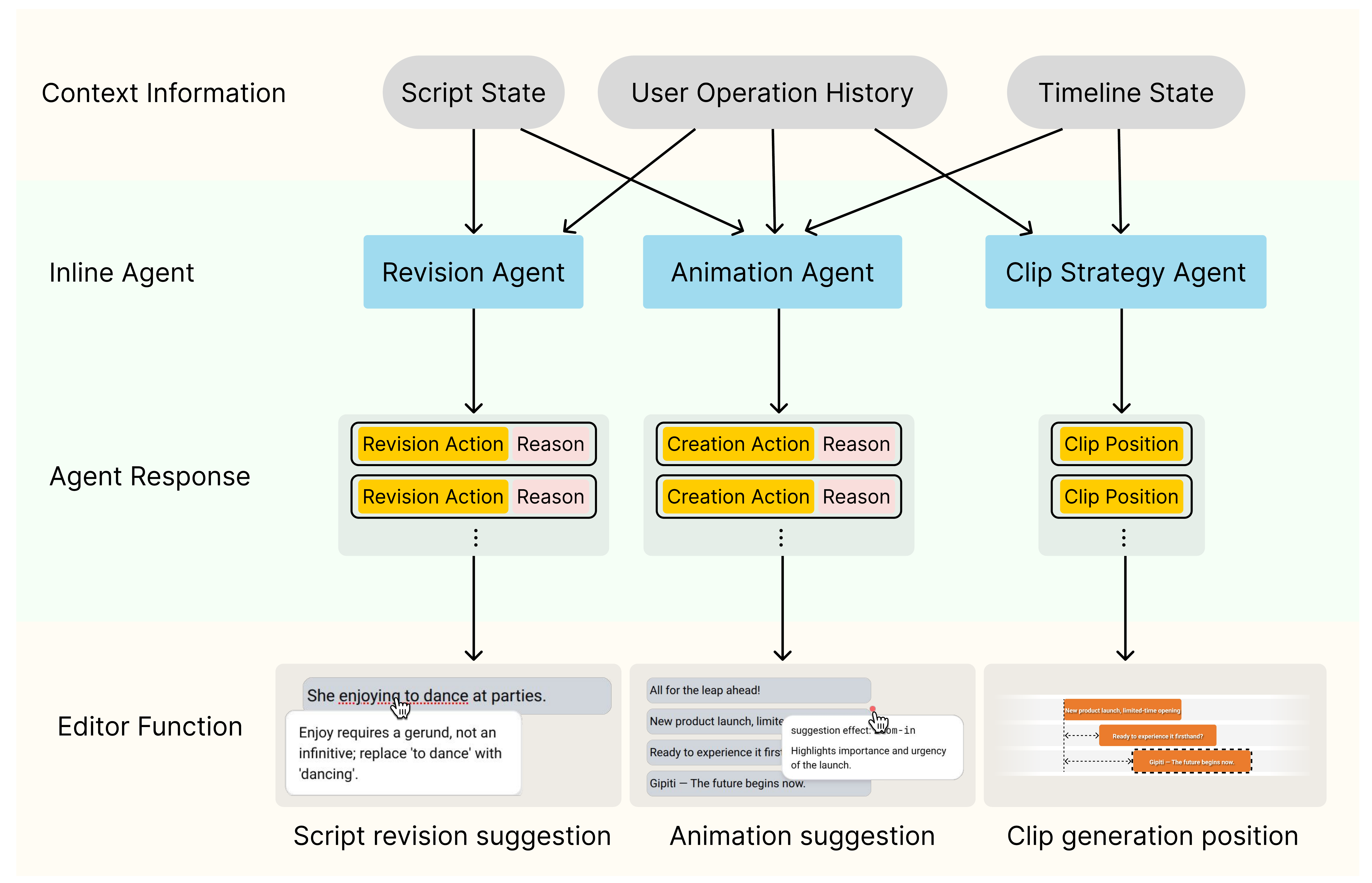}
\caption{Inline Agents Design: Different editing context information is routed to specialized inline agents as input. 
Each agent is designed to perform a single, well-defined task.
All agent responses are structured by design, and their execution, upon user approval, is handled by the editor's internal functions.
}
\vspace{-0.46cm}
\label{fig:inline_agent}
\end{figure}

\subsubsection{Inline Agent (\autoref{fig:inline_agent})}
We design Inline Agents to support lightweight, context-specific interactions within the editor. Different editing context information is routed to specialized inline agents as input, where each agent is responsible for performing a single, well-defined task, such as suggesting text edits or recommending animations. 

All agent responses are structured, providing both the proposed action and an explanatory rationale to help users understand the suggestion. Upon user approval, the corresponding actions are executed through the editor's internal functions. This design allows users to maintain control over the editing process while benefiting from intelligent, task-specific assistance without disrupting their workflow.

For suggestion-generating agents, they are specifically designed to additionally generate a reason alongside each action. This reason serves as a description of the suggested action, ensuring that users have a clear understanding of the
proposed changes before they are applied. For Clip strategy agent, it is tasked with generating a new clip position based on predefined strategy and context information, which include both track and start time of clip.

\begin{figure}[h!]
\centering
\includegraphics[width=0.5\textwidth]{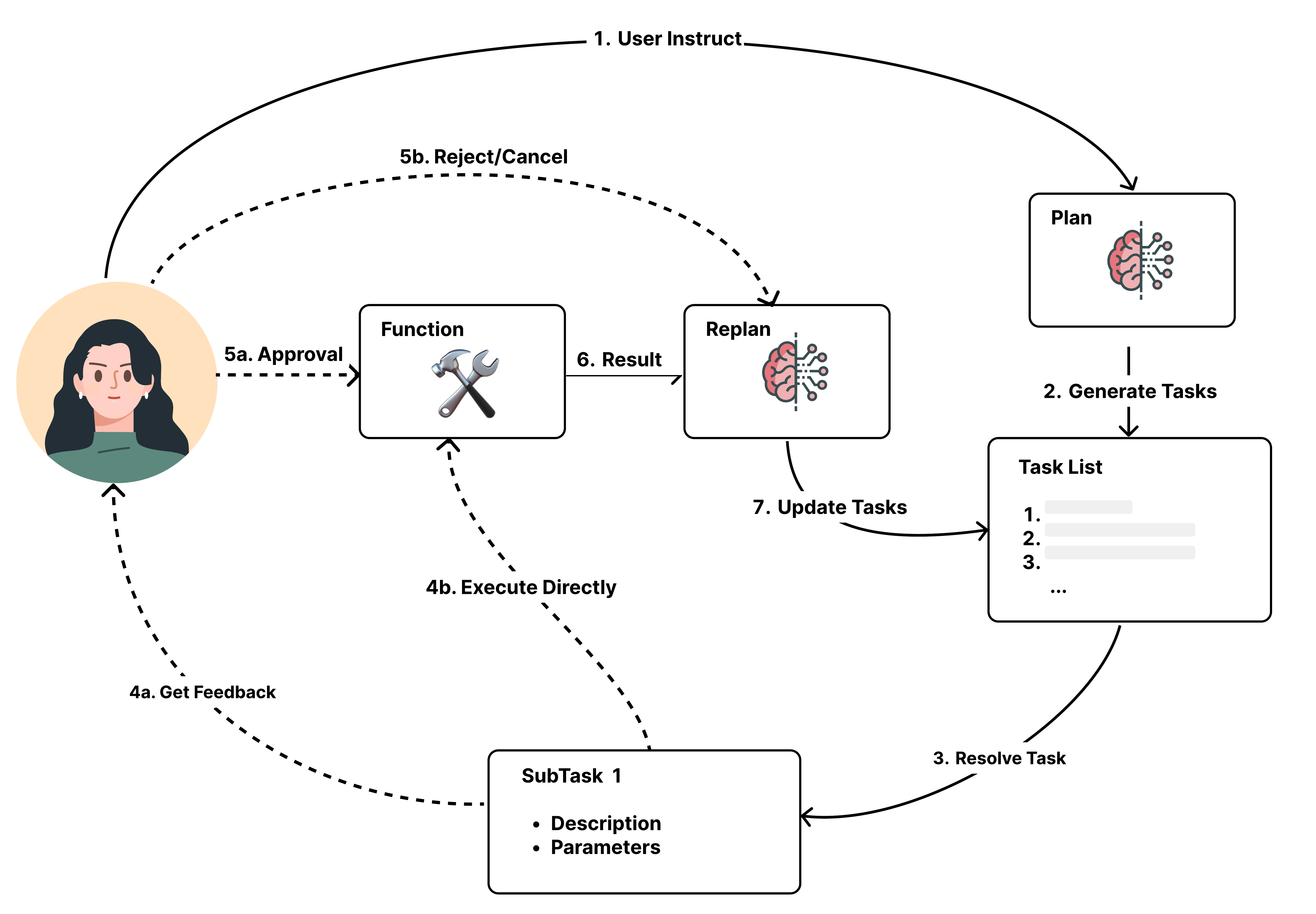}
\caption{Chat Agent Loop: The Chat Agent follows a Plan-and-Execute style. 
For editing tasks, functions are executed upon user approval, while for query tasks, execution occurs automatically. 
The agent updates the task list based on execution results and user feedback, continuing until all tasks are completed.
}
\label{fig:chat-agent}
\end{figure}

\subsubsection{Chat Agent (\autoref{fig:chat-agent})}
The Chat Agent leverages large language models (LLMs) to perform advanced reasoning and task execution through function calls. 
It is designed to facilitate natural language interactions, enabling users to control and manipulate various editable objects within the system. 
By utilizing LLMs, the Chat Agent can interpret user commands and trigger corresponding actions, such as creating, modifying, or querying objects in the editing environment. 
This approach enhances the flexibility of interactions and allows users to perform complex tasks with minimal effort and without requiring specific technical knowledge.

The Chat Agent use a Plan-and-Execute\cite{Zhao:2024:lightva,Myers:1999:cpef,Zelong:2024:formalllm,Despouys:2000:propice} (Figure. \ref{fig:chat-agent}) pattern to provide assistance to users. When users input editing objectives, the agent responds through a multi-step process: 
\begin{itemize}
    \item \textbf{Plan}: The agent parses the user's editing objective and generates a one-step execution plan. Users can modify parameters in the execution plan or reject the plan;
    \item \textbf{Execute}: After user approval, the agent executes the planned actions and analyzes whether a subsequent execution plan is needed. If required, it re-enters the planning phase; if not, it outputs a text prompt and concludes the current editing session.
\end{itemize}
The chat agent includes an auto-skip option. When selected, the current editing session automatically approves all system editing operations without requiring multiple user approvals.

\subsection{Implementation of LLM-Powered Editing Functions}
The Agent component of this system provides several core functionalities: 1) Editing intent comprehension; 2) Semantic-animation matching; 3) Video element modification; 4) Text content refinement. All these functionalities are constructed based on LLM prompt engineering, with their operations grounded in real-time monitoring of various editor states. Below, we will first elaborate on our editor state monitoring methodology, followed by detailed explanations of each functional implementation.

In common editor systems (e.g., game editor, video editor), meta-objects are commonly used to represent objects, encompassing the type, value range, description, and tooltip for each attribute.
Meta-objects typically support factory-based object creation, serialization and deserialization, and parameter panel reflection within editor architectures.

In our system, we similarly introduce a meta-object framework into the Agent implementation.
This approach enables automatic generation of Function Calling parameters and provides a unified representation of editable objects and editor state, significantly enhancing the flexibility and extensibility of the system.

\subsubsection{Editor State Monitoring}
In our text animation editor, we implement a meta-object-based state and tool management system. This architecture enables simple addition of new element types and tools without modifying Agent prompts, significantly enhancing system extensibility and providing a reference paradigm for future AI-assisted editing system development.
\begin{itemize}
    \item \textbf{Engineering Content Management}: The text animation system is built upon a meta-object framework. We define three meta-object classes: Assets, Timeline Elements, and Animation Effects. All materials and timeline elements within the editor are instances of these classes. Each instance possesses a unique ID for precise reference during LLM-driven modification tasks. All meta-object classes and instances maintain direct reflection relationships with UI components, thereby simplifying UI design and update processes.
    \item \textbf{Tool Management}: For meta-object classes requiring LLM operations, we design dedicated manipulation functions. Each meta-object operation supports both individual and batch processing modes to accommodate precise and bulk editing requirements. System-defined functions can be directly comprehended and invoked by LLMs without prompt modifications.
\end{itemize}

\subsubsection{Editing Intent Comprehension}
To better align with user expectations, we employ specialized prompt templates requiring the Agent to comprehend user editing intent through accessible inputs before providing assistance. The Agent processes: user dialog inputs, current timeline elements, text content, operation logs, and material resources. These inputs enable the LLM to interpret user intentions and reference contextual information when generating assistive operations.

\subsubsection{Semantic-Animation Matching}
For text-driven animation recommendation tasks, we establish a semantic-animation mapping framework. The LLM Agent utilizes this mapping through specifically designed prompt templates to provide contextually appropriate animation suggestions that align with textual content.

\subsubsection{Video Element Modification}
When executing video element modification tasks, the LLM initially aggregates user dialog inputs, current timeline elements, text content, material resources, and operation logs. By analyzing creative intent, the LLM generates corresponding tool invocations, iteratively producing subsequent operations after user approval. The prompt template terminates the modification process when no further tool invocations are required, accompanied by user feedback prompts.

\subsubsection{Text Content Refinement}
Leveraging LLM's linguistic capabilities during text refinement tasks, the system sequentially generates tool invocations for modifying individual text units. Each textual modification requires user approval, with the process concluding through prompt template-driven user notifications upon task completion.

\subsection{System Implementation}
We implemented our system as a full-stack web application. For LLm inference, we incorporated the latest GPT-4o model from OpenAI to power the agents. The front-end is built with React.js, integrating Motion-Canvas\footnote{\url{https://motioncanvas.io/}} for animation editing. For improved workflow, particularly in Script-Timeline synchronization editing, Slate.js\footnote{\url{https://docs.slatejs.org/}} is utilized to manage reference and linkage. The agents are constructed using LangGraph\footnote{\url{https://www.langchain.com/langgraph}} in combination with Copilot-Kit\footnote{\url{https://www.copilotkit.ai/}}, facilitating robust natural language processing and interaction. This architecture ensures seamless collaboration and editing capabilities throughout the system.

\section{User Study}
We conducted a user study to gather feedback on users' experience with the text animation editing system. The research objectives include: 1) evaluating the effectiveness of the LLM-powered agent functionality in the text animation editing system for assisting with text animation editing tasks; 2) understanding users' perception of the LLM-based editing assistant agent during the editing process. We invited participants to create text animation content with both assigned themes and free themes using the system, in order to test the system's functionality and practicality across different creative tasks. In presenting the results, we ralate the findings to the design goals of the system (\textbf{DG1, DG2, DG3}), highlighting their fulfillment.

\subsection{Participants}
We focused on users with varying levels of video editing experience and their acceptance of LLM-driven text animation-assisted editing. To this end, we recruited 11 participants from diverse backgrounds (3 female, mean = 21, SD = 2.24). Their video editing experience ranged from 1 to 8 years (mean = 3.45, SD = 2.16), and their specific backgrounds are shown in the table. Some participants frequently create videos: P1 and P3 produce promotional videos for student clubs and edit daily vlogs; P5 creates knowledge-sharing videos for self-media platforms; P6 and P10 work on promotional videos for commercial products; and P4 produces anime-style compilation videos. All participants had prior experience using LLM tools but had never used conversational LLM tools integrated with video editing.

\begin{table}[h!]
\centering
\begin{tabular}{|c|c|c|c|c|}
\hline
\textbf{Participant} & \textbf{Age} & \textbf{Gender} & \textbf{Identity} & \textbf{Years of Experience} \\ \hline
P1 & 19 & Male & Student & 4 \\ \hline
P2 & 18 & Male & Student & 3 \\ \hline
P3 & 20 & Male & Student & 3 \\ \hline
P4 & 19 & Male & Student & 6 \\ \hline
P5 & 24 & Male & Student & 4 \\ \hline
P6 & 25 & Female & Staff & 1 \\ \hline
P7 & 20 & Male & Student & 1 \\ \hline
P8 & 23 & Female & Student & 1 \\ \hline
P9 & 20 & Male & Student & 8 \\ \hline
P10 & 22 & Female & Staff & 4 \\ \hline
P11 & 21 & Male & Student & 3 \\ \hline
\end{tabular}
\caption{Participant demographics and experience in user study.}
\label{tab:participants}
\end{table}
\subsection{Procedure}
\begin{figure}[h!]
\centering
\includegraphics[width=0.5\textwidth]{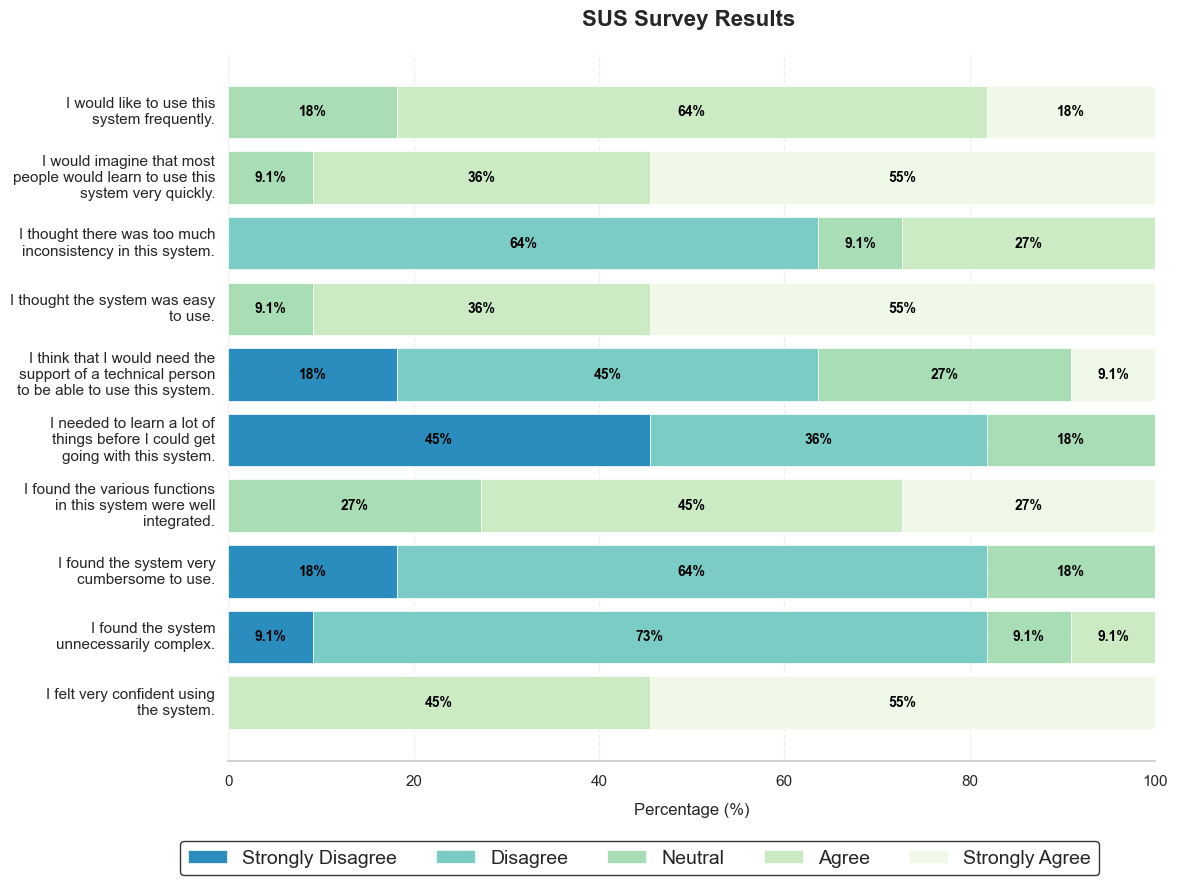}
\caption{Distribution of SUS usability scale results for the text animation editing system. Overall, participants held positive attitudes toward their experience with the text animation editing system and its usability evaluation, though they expressed some reservations on certain items.}
\label{fig:sus}
\end{figure}

The study was conducted in a quiet environment and lasted approximately 1 to 1.5 hours. Participants first received a 10-15 minute introduction to the system's functionalities, then used the text animation editing system to complete two tasks. After completing the tasks, they filled out a questionnaire to evaluate the system's usability, particularly focusing on the LLM-powered agent, using the SUS (System Usability Scale) and the results shown in~\autoref{fig:sus}. Finally, they participated in a 10-15 minute semi-structured interview. Throughout the process, participants were encouraged to ask questions and share their opinions, with a focus on their genuine user experience.

The two tasks completed by participants were as follows:

\begin{itemize}
    \item First, we provided a specified text passage and asked users to edit it into a text animation both with and without the assistance of the LLM-based editing tool.
    \item Then, we asked users to freely create text animations using the system, constructing forms they might use in their regular workflows, and to compare this experience with their usual workflow.
\end{itemize}

\subsection{Results}

\subsubsection{Editing Effectiveness and Overall Impressions}
All users quickly mastered the system's operation during the study (mean learning cost = 1.73, SD = 0.79). The 11 participants gave the system an SUS usability score of 75. Users found the text animation editing system easy to use (mean = 4.45, SD = 0.69) and expressed willingness to use it frequently (mean = 4, SD = 0.63). Amateur video creators, such as P11, were pleasantly surprised by the system's capabilities: ``\textit{The agent can control many aspects to complete editing.}''

The study results highlight the text animation editing system's strengths in providing a low learning curve and an intuitive human-computer interaction design of the agent-based pipeline. Participants quickly adapted to the system, demonstrating its accessibility even for users with limited technical expertise. The conversational interface, powered by the LLM-powered agent, was particularly praised for its ability to simplify complex editing tasks and streamline workflows. This design not only reduced the cognitive load on users but also enhanced their overall editing efficiency. The system's usability score of 75 further underscores its effectiveness in meeting user needs, with participants expressing satisfaction with its ease of use and functionality. These findings suggest that the system successfully balances technical sophistication with user-centered design principles, making it a practical tool for both amateur and professional video creators. This further validates the effectiveness of the human-computer interaction paradigm adopted in our proposed pipeline.

\subsubsection{Comparison with Existing Tools}
Users acknowledged the innovative paradigm of the text animation editor in the video editing domain. P1, who is familiar with video editing tools, remarked: ``\textit{Currently, there are no video editing tools on the market that adjust parameters through conversational interaction.}'' P4 likened the LLM-powered agent to an assistant added to existing video editors, specifically handling tedious operations:``\textit{With the agent handling these details, I can focus more on adjusting the visual effects.}'' P5, who has experience with LLM tools, noted that the system's conversational interaction is more efficient than traditional video editing software. P7, who is less familiar with video editing, found the system's conversational interaction more intuitive than traditional software.

In summary, the text animation editor introduces a novel approach to video editing by leveraging conversational interaction, which users found both efficient and intuitive compared to traditional video editing tools. Unlike conventional editors that often require extensive manual adjustments, this system streamlines tedious tasks and simplifies workflows through its LLM-powered conversational interface. Experienced users highlighted its ability to reduce repetitive operations, while less experienced users appreciated its accessibility and ease of use. The integration of an agent as a conversational assistant was particularly praised for offering a more intuitive and efficient alternative to traditional parameter-based editing. These findings demonstrate the system's potential to redefine video editing practices by providing a user-friendly and innovative solution that surpasses the limitations of traditional tools.

\subsubsection{Usability and Practicality}
Overall, users gave positive feedback on the text animation editing system, though opinions diverged on certain specific features. Negative feedback primarily stemmed from the following scenarios: professional users (e.g., P6) desired more advanced editing capabilities, and LLM API performance issues resulted in the agent's response time and randomness not fully meeting usability expectations. Specific feedback included:

\begin{itemize}
    \item \textbf{Draft creation}: P4 noted that directly generating a draft project eliminated the need for manual animation editing, saving time. P5 appreciated the convenience of the conversational agent generating drafts from existing materials but wished different operations could be performed concurrently to save time.
    \item \textbf{Text refinement}: All users praised the LLM's ability to refine text content. P11 believed this feature made it easier to emulate different styles.
    \item \textbf{Batch modifications}: Most users found batch execution of simple operations practical for video editing tasks (P4, P11), though P6 hoped the agent could support advanced editing techniques.
\end{itemize}

Additionally, users suggested potential system improvements. P6 proposed that the agent could implement existing editing techniques, such as varied mask applications. P9 suggested the current agent could serve as a plugin for mature editors like Premiere Pro, combining LLM comprehension with robust editing features. P2 and P4 recommended adding internet connectivity and multimodal capabilities to enable direct online material searches and insertion into video projects.

In summary, users expressed strong appreciation for the system's three core functionalities: draft creation, text refinement, and batch modifications. The draft creation feature was particularly valued for its ability to save time by automating the initial stages of animation editing, allowing users to focus on creative adjustments. Text refinement, powered by the LLM, received widespread praise for its ability to enhance content quality and emulate diverse styles, making it a standout feature for users aiming for polished results. Batch modifications were highlighted as a practical tool for streamlining repetitive tasks, especially for users managing large-scale projects. These features collectively enhanced the system's usability and practicality, making it an effective tool for both amateur and professional video creators.

Despite these strengths, users also identified areas for improvement. Professional users suggested incorporating advanced editing capabilities, such as multi-layer editing and real-time collaboration, to better meet their needs. Additionally, addressing LLM API performance issues, such as response time and occasional randomness, could further enhance the system's reliability. Suggestions like integrating the agent as a plugin for mature editors and adding internet connectivity for online material searches also point to potential directions for future development. These insights highlight the system's current success while providing a roadmap for continued refinement and innovation.

\subsubsection{Workflow Innovation}
Most users agreed the system could optimize their current workflows. It helped construct initial video drafts, saving time. P1 stated: ``\textit{With the script already prepared, the agent can directly assemble timeline elements.}'' The conversational editing feature also enabled batch modifications, reducing workload. P4 remarked: ``\textit{Compared to before, I can now batch-apply animations first and make minor adjustments later.}''

The text animation editing system significantly improved users' workflows by addressing common pain points in video editing. By automating repetitive tasks, such as constructing initial drafts and enabling batch modifications, the system reduced the time and effort required for manual adjustments. Users noted that these features allowed them to focus more on creative refinements, streamlining their overall editing process. The conversational interface, powered by LLMs, further enhanced efficiency by simplifying complex operations and reducing cognitive load. While the system effectively optimized workflows, users also identified areas for improvement, such as incorporating advanced editing capabilities and addressing LLM performance issues. These insights highlight the system's impact on improving existing workflows while pointing to opportunities for further refinement.

\subsubsection{LLM Utilization Experience}
Since the system relies on LLMs to power the agent-assisted editing, users' familiarity with LLM characteristics influenced collaboration effectiveness. P5 observed: ``\textit{If you're already skilled with LLM tools, the editing results tend to be better.}'' This observation highlights the importance of user expertise in maximizing the system's potential. Participants with prior experience using LLM tools were able to leverage the agent more effectively, utilizing advanced prompts and commands to achieve precise editing outcomes. For example, P3 noted that understanding how to phrase requests allowed them to generate more accurate animations, while P11 emphasized the value of experimenting with different input styles to refine results. Conversely, less experienced users, such as P8, mentioned a steeper learning curve when interacting with the agent, requiring additional time to understand its capabilities. These findings suggest that while the system offers significant benefits, providing training or guidance on effective LLM usage could further enhance user experience and reduce the gap between novice and experienced users.

\section{Design Implication}
Based on the findings of the previous sections, we discuss implications for the design of the future LLM-based assistances.

\subsection{Designing Multiple Agents in Editors to Adapt to Different Interaction Forms}
This study demonstrates the effectiveness of employing multiple agents in content editors to assist with editing tasks in different content areas. In traditional editors, different types of content often require distinct editing forms to accommodate their characteristics. Designing agents that adapt to these traditional editing forms in different content areas can align with users' habits, reduce their learning costs, and achieve a higher level of human-computer interaction. This approach is not limited to text-animation editors but can also be applied to various content editing domains, such as video editing and game engines, to design AI-assisted editing agents.

\subsection{Using UI Prompts to Handle Ambiguous Commands}
Due to the limitations of LLM performance or ambiguities in user expressions\cite{Hyuhng:2024:aligning, Sanidhya:2025:interactive, Park:2024:clara}, LLM agents in various assistive editing systems may sometimes fail to identify the elements referred to by users. To address this issue, this study proposes a human-computer interaction-based solution. When the LLM detects ambiguity, it is designed not to directly execute operations but instead to output a UI component prompting\cite{Stephen:2023:prompt,Torricelli:2024:theRole,Strobelt:2023:interactive} the user to specify the element to be operated on. Using UI prompts effectively reduces the demands on users' linguistic expression capabilities. Feeding this information back into the LLM as a new prompt can prevent further ambiguities. Therefore, we recommend that future LLM applications incorporate ambiguity detection and, when the LLM cannot determine the user's intent, present the issue to the user in the form of a UI prompt. This design can enhance the agent's accuracy and achieve better human-computer interaction outcomes.

\subsection{Allowing Users to Precisely Locate Elements for Agents}
This study shows that allowing users to precisely locate elements in LLM-based human-computer interaction systems can improve user experience. Direct manipulation through drag-and-drop referencing is more intuitive for users and has a lower learning cost, while referencing via special symbols avoids attention shifts, enabling more experienced users to operate quickly. Precise element location allows users to express their operational intentions more conveniently, while the LLM can better understand these intentions, resulting in more efficient and accurate human-computer interaction. Future systems integrating LLM agents could consider incorporating methods to facilitate precise element location, such as providing options for selecting interface elements.

\subsection{Combining AI Agents with Meta-Objects}
Meta-objects are widely used in content editors\cite{CapCut, AdobePremierePro}, such as game engines and video editors. This approach enables direct mapping from classes, instances, and methods to parameter panels, facilitating the integration of content descriptions with the UI. Combining AI agents with meta-objects can reduce the number of prompts and interfaces required by AI agents. Additionally, when new features are added to the editor, there is no need to modify the prompts, significantly enhancing the system's scalability. For future assistive editing tools that need to be implemented as plugins in other editors, the values of meta-objects in the editor can be directly utilized to achieve highly scalable AI-assisted tools, reducing system development and maintenance costs.

\section{Limitations and Future Work}
The current system is an attempt to explore the emerging field of LLM-assisted content editing system design. Given the rapid development of LLM research, the current design and implementation are time-sensitive. We believe the enduring value of this work lies not in the specific implementations that may undergo rapid iterations, but in its pioneering exploration of LLM editing for specific animation effects and parameters, which lays the foundation for the continued development of this field. Below, we will discuss the limitations of our current system and future work.

\subsection{Timeline-Script Collaborative Editing}
In this system, we designed bidirectional synchronization between the timeline and text content editing, allowing users to control certain timeline properties within the script editing panel. This approach enhances system consistency and reduces the need for users to shift their attention. However, the elements in the current script editing panel remain limited, and some timeline properties and specific effects of text animations cannot be represented or modified in the script editing panel. In the future, more visual encodings could be added to represent additional timeline properties in the script editing panel, and more reasonable visual designs could be employed to display specific animation effects in the script editing interface.

\subsection{Agent Architecture}
The current Chat agent is implemented as a single-agent system. To enable the agent to assist with a wide range of editing tasks, it is bound to all the functions supported by the system. This results in significant token consumption. In the future, a multi-agent design could be introduced, where different tasks are assigned to specialized agents. This would reduce token consumption without compromising the agent’s intelligence, thereby improving its cost-effectiveness.

Additionally, the LLM currently employed in the system is a text-only model, which has limited capacity to understand specific visual effects and asset content. In the future, multimodal large models could be integrated into the system, allowing for a deeper understanding of the animation effects and visual elements within video projects.

\subsection{Research Domain}
This system focuses on the editing of text animations. However, during user research, we found that its workflow could also be applied to other content editing domains, such as broader video editing tasks. In the future, the work presented here could be extended to the field of video editing through more reasonable agent and interaction designs, providing editing assistance to a wider range of video creators.

\subsection{User Evaluation}
Our user study evaluated the text animation editing system with 11 participants from diverse backgrounds to obtain multifaceted perspectives. However, the sample size may affect the generalizability of the conclusions. Future research could expand the number of participants, covering more diverse user backgrounds and different video editing scenarios, to validate and extend the preliminary findings presented here. Additionally, future work may involve inviting professionals from the film and television industry for user evaluations, providing more specialized assessments of future tools with enhanced performance.

\section{Conclusion}
We present Text Animation Editing System, a text-based animation editing tool that realizes an agent-assisted animation parameter editing paradigm through LLM-based agent assistance. We detail the system's unique design and implementation solutions, including its user interface modules and backend architectural execution. Through user studies, we evaluated the practical efficacy of the Text Animation Editing System and gained insights into users' perceptions and feedback regarding LLM-based agent involvement in animation editing workflows. Based on research outcomes, we propose academic implications to guide future development of LLM-assisted content editing systems. This work illuminates potential directions for agent-assisted media content editing tools, and we anticipate the emergence of more sophisticated auxiliary editing tools in the future.

\begin{acks}
To Robert, for the bagels and explaining CMYK and color spaces.
\end{acks}

\bibliographystyle{ACM-Reference-Format}
\bibliography{sample-base}


\begin{thebibliography}{57}


\ifx \showCODEN    \undefined \def \showCODEN     #1{\unskip}     \fi
\ifx \showISBNx    \undefined \def \showISBNx     #1{\unskip}     \fi
\ifx \showISBNxiii \undefined \def \showISBNxiii  #1{\unskip}     \fi
\ifx \showISSN     \undefined \def \showISSN      #1{\unskip}     \fi
\ifx \showLCCN     \undefined \def \showLCCN      #1{\unskip}     \fi
\ifx \shownote     \undefined \def \shownote      #1{#1}          \fi
\ifx \showarticletitle \undefined \def \showarticletitle #1{#1}   \fi
\ifx \showURL      \undefined \def \showURL       {\relax}        \fi
\providecommand\bibfield[2]{#2}
\providecommand\bibinfo[2]{#2}
\providecommand\natexlab[1]{#1}
\providecommand\showeprint[2][]{arXiv:#2}

\bibitem[AI({[n.\,d.]})]%
        {StabilityAI}
\bibfield{author}{\bibinfo{person}{Stability AI}.} \bibinfo{year}{[n.\,d.]}\natexlab{}.
\newblock \bibinfo{booktitle}{\emph{Stability AI - The Foundation of Generative AI}}.
\newblock
\urldef\tempurl%
\url{https://stability.ai/}
\showURL{%
\tempurl}


\bibitem[Amini et~al\mbox{.}(2015)]%
        {10.1145/2702123.2702431}
\bibfield{author}{\bibinfo{person}{Fereshteh Amini}, \bibinfo{person}{Nathalie Henry~Riche}, \bibinfo{person}{Bongshin Lee}, \bibinfo{person}{Christophe Hurter}, {and} \bibinfo{person}{Pourang Irani}.} \bibinfo{year}{2015}\natexlab{}.
\newblock \showarticletitle{Understanding Data Videos: Looking at Narrative Visualization through the Cinematography Lens}. In \bibinfo{booktitle}{\emph{Proceedings of the 33rd Annual ACM Conference on Human Factors in Computing Systems}} (Seoul, Republic of Korea) \emph{(\bibinfo{series}{CHI '15})}. \bibinfo{publisher}{Association for Computing Machinery}, \bibinfo{address}{New York, NY, USA}, \bibinfo{pages}{1459–1468}.
\newblock
\showISBNx{9781450331456}
\href{https://doi.org/10.1145/2702123.2702431}{doi:\nolinkurl{10.1145/2702123.2702431}}


\bibitem[Barua et~al\mbox{.}(2025)]%
        {10.1145/3708359.3712090}
\bibfield{author}{\bibinfo{person}{Aadit Barua}, \bibinfo{person}{Karim Benharrak}, \bibinfo{person}{Meng Chen}, \bibinfo{person}{Mina Huh}, {and} \bibinfo{person}{Amy Pavel}.} \bibinfo{year}{2025}\natexlab{}.
\newblock \showarticletitle{Lotus: Creating Short Videos From Long Videos With Abstractive and Extractive Summarization}. In \bibinfo{booktitle}{\emph{Proceedings of the 30th International Conference on Intelligent User Interfaces}} \emph{(\bibinfo{series}{IUI '25})}. \bibinfo{publisher}{Association for Computing Machinery}, \bibinfo{address}{New York, NY, USA}, \bibinfo{pages}{967–981}.
\newblock
\showISBNx{9798400713064}
\href{https://doi.org/10.1145/3708359.3712090}{doi:\nolinkurl{10.1145/3708359.3712090}}


\bibitem[{ByteDance Ltd.}({[n.\,d.]})]%
        {CapCut}
\bibfield{author}{\bibinfo{person}{{ByteDance Ltd.}}} \bibinfo{year}{[n.\,d.]}\natexlab{}.
\newblock \bibinfo{booktitle}{\emph{CapCut: Free Video Editing Software}}.
\newblock
\urldef\tempurl%
\url{https://www.capcut.com/}
\showURL{%
\tempurl}


\bibitem[Cabrera et~al\mbox{.}(2023)]%
        {10.1145/3579612}
\bibfield{author}{\bibinfo{person}{\'{A}ngel~Alexander Cabrera}, \bibinfo{person}{Adam Perer}, {and} \bibinfo{person}{Jason~I. Hong}.} \bibinfo{year}{2023}\natexlab{}.
\newblock \showarticletitle{Improving Human-AI Collaboration With Descriptions of AI Behavior}.
\newblock \bibinfo{journal}{\emph{Proc. ACM Hum.-Comput. Interact.}} \bibinfo{volume}{7}, \bibinfo{number}{CSCW1}, Article \bibinfo{articleno}{136} (\bibinfo{date}{April} \bibinfo{year}{2023}), \bibinfo{numpages}{21}~pages.
\newblock
\href{https://doi.org/10.1145/3579612}{doi:\nolinkurl{10.1145/3579612}}


\bibitem[Cardelli(1988)]%
        {10.1145/62402.62428}
\bibfield{author}{\bibinfo{person}{Luca Cardelli}.} \bibinfo{year}{1988}\natexlab{}.
\newblock \showarticletitle{Building user interfaces by direct manipulation}. In \bibinfo{booktitle}{\emph{Proceedings of the 1st Annual ACM SIGGRAPH Symposium on User Interface Software}} (Alberta, Canada) \emph{(\bibinfo{series}{UIST '88})}. \bibinfo{publisher}{Association for Computing Machinery}, \bibinfo{address}{New York, NY, USA}, \bibinfo{pages}{152–166}.
\newblock
\showISBNx{0897912837}
\href{https://doi.org/10.1145/62402.62428}{doi:\nolinkurl{10.1145/62402.62428}}


\bibitem[Ceylan et~al\mbox{.}(2023)]%
        {10.48550/arXiv.2303.12688}
\bibfield{author}{\bibinfo{person}{Duygu Ceylan}, \bibinfo{person}{Chun-Hao~Paul Huang}, {and} \bibinfo{person}{Niloy~J. Mitra}.} \bibinfo{year}{2023}\natexlab{}.
\newblock \bibinfo{title}{Pix2Video: Video Editing using Image Diffusion}.
\newblock
\href{https://doi.org/10.48550/arXiv.2303.12688}{doi:\nolinkurl{10.48550/arXiv.2303.12688}}
\showeprint[arxiv]{2303.12688}~[cs.CV]


\bibitem[Cohen et~al\mbox{.}(1989)]%
        {10.1145/67450.67494}
\bibfield{author}{\bibinfo{person}{P.~R. Cohen}, \bibinfo{person}{M. Dalrymple}, \bibinfo{person}{D.~B. Moran}, \bibinfo{person}{F.~C. Pereira}, {and} \bibinfo{person}{J.~W. Sullivan}.} \bibinfo{year}{1989}\natexlab{}.
\newblock \showarticletitle{Synergistic use of direct manipulation and natural language}.
\newblock \bibinfo{journal}{\emph{SIGCHI Bull.}} \bibinfo{volume}{20}, \bibinfo{number}{SI} (\bibinfo{date}{March} \bibinfo{year}{1989}), \bibinfo{pages}{227–233}.
\newblock
\showISSN{0736-6906}
\href{https://doi.org/10.1145/67450.67494}{doi:\nolinkurl{10.1145/67450.67494}}


\bibitem[Croitoru et~al\mbox{.}(2023)]%
        {10.1109/TPAMI.2023.3261988}
\bibfield{author}{\bibinfo{person}{Florinel-Alin Croitoru}, \bibinfo{person}{Vlad Hondru}, \bibinfo{person}{Radu~Tudor Ionescu}, {and} \bibinfo{person}{Mubarak Shah}.} \bibinfo{year}{2023}\natexlab{}.
\newblock \showarticletitle{Diffusion Models in Vision: A Survey}.
\newblock \bibinfo{journal}{\emph{IEEE Transactions on Pattern Analysis and Machine Intelligence}} \bibinfo{volume}{45}, \bibinfo{number}{9} (\bibinfo{year}{2023}), \bibinfo{pages}{10850--10869}.
\newblock
\href{https://doi.org/10.1109/TPAMI.2023.3261988}{doi:\nolinkurl{10.1109/TPAMI.2023.3261988}}


\bibitem[Despouys and Ingrand(2000)]%
        {Despouys:2000:propice}
\bibfield{author}{\bibinfo{person}{Olivier Despouys} {and} \bibinfo{person}{Fran{\c{c}}ois~F{\'e}lix Ingrand}.} \bibinfo{year}{2000}\natexlab{}.
\newblock \showarticletitle{Propice-Plan: Toward a Unified Framework for Planning and Execution}. In \bibinfo{booktitle}{\emph{Recent Advances in AI Planning}}, \bibfield{editor}{\bibinfo{person}{Susanne Biundo} {and} \bibinfo{person}{Maria Fox}} (Eds.). \bibinfo{publisher}{Springer Berlin Heidelberg}, \bibinfo{address}{Berlin, Heidelberg}, \bibinfo{pages}{278--293}.
\newblock
\showISBNx{978-3-540-44657-6}


\bibitem[Drori and Te'eni(2024)]%
        {10.17705/1jais.00867}
\bibfield{author}{\bibinfo{person}{Iddo Drori} {and} \bibinfo{person}{Dov Te'eni}.} \bibinfo{year}{2024}\natexlab{}.
\newblock \showarticletitle{Human-in-the-Loop AI Reviewing: Feasibility, Opportunities, and Risks}.
\newblock \bibinfo{journal}{\emph{Journal of the Association for Information Systems}}  \bibinfo{volume}{25} (\bibinfo{date}{01} \bibinfo{year}{2024}), \bibinfo{pages}{98--109}.
\newblock
\href{https://doi.org/10.17705/1jais.00867}{doi:\nolinkurl{10.17705/1jais.00867}}


\bibitem[Fan et~al\mbox{.}(2024)]%
        {10.1145/3613904.3642129}
\bibfield{author}{\bibinfo{person}{Xianzhe Fan}, \bibinfo{person}{Zihan Wu}, \bibinfo{person}{Chun Yu}, \bibinfo{person}{Fenggui Rao}, \bibinfo{person}{Weinan Shi}, {and} \bibinfo{person}{Teng Tu}.} \bibinfo{year}{2024}\natexlab{}.
\newblock \showarticletitle{ContextCam: Bridging Context Awareness with Creative Human-AI Image Co-Creation}. In \bibinfo{booktitle}{\emph{Proceedings of the 2024 CHI Conference on Human Factors in Computing Systems}} (Honolulu, HI, USA) \emph{(\bibinfo{series}{CHI '24})}. \bibinfo{publisher}{Association for Computing Machinery}, \bibinfo{address}{New York, NY, USA}, Article \bibinfo{articleno}{157}, \bibinfo{numpages}{17}~pages.
\newblock
\showISBNx{9798400703300}
\href{https://doi.org/10.1145/3613904.3642129}{doi:\nolinkurl{10.1145/3613904.3642129}}


\bibitem[Fried et~al\mbox{.}(2019)]%
        {10.1145/3306346.3323028}
\bibfield{author}{\bibinfo{person}{Ohad Fried}, \bibinfo{person}{Ayush Tewari}, \bibinfo{person}{Michael Zollh\"{o}fer}, \bibinfo{person}{Adam Finkelstein}, \bibinfo{person}{Eli Shechtman}, \bibinfo{person}{Dan~B Goldman}, \bibinfo{person}{Kyle Genova}, \bibinfo{person}{Zeyu Jin}, \bibinfo{person}{Christian Theobalt}, {and} \bibinfo{person}{Maneesh Agrawala}.} \bibinfo{year}{2019}\natexlab{}.
\newblock \showarticletitle{Text-based editing of talking-head video}.
\newblock \bibinfo{journal}{\emph{ACM Trans. Graph.}} \bibinfo{volume}{38}, \bibinfo{number}{4}, Article \bibinfo{articleno}{68} (\bibinfo{date}{July} \bibinfo{year}{2019}), \bibinfo{numpages}{14}~pages.
\newblock
\showISSN{0730-0301}
\href{https://doi.org/10.1145/3306346.3323028}{doi:\nolinkurl{10.1145/3306346.3323028}}


\bibitem[Hitsuwari et~al\mbox{.}(2023)]%
        {10.1016/j.chb.2022.107502}
\bibfield{author}{\bibinfo{person}{Jimpei Hitsuwari}, \bibinfo{person}{Yoshiyuki Ueda}, \bibinfo{person}{Woojin Yun}, {and} \bibinfo{person}{Michio Nomura}.} \bibinfo{year}{2023}\natexlab{}.
\newblock \showarticletitle{Does human–AI collaboration lead to more creative art? Aesthetic evaluation of human-made and AI-generated haiku poetry}.
\newblock \bibinfo{journal}{\emph{Computers in Human Behavior}}  \bibinfo{volume}{139} (\bibinfo{year}{2023}), \bibinfo{pages}{107502}.
\newblock
\showISSN{0747-5632}
\href{https://doi.org/10.1016/j.chb.2022.107502}{doi:\nolinkurl{10.1016/j.chb.2022.107502}}


\bibitem[Huber et~al\mbox{.}(2019)]%
        {10.1145/3290605.3300311}
\bibfield{author}{\bibinfo{person}{Bernd Huber}, \bibinfo{person}{Hijung~Valentina Shin}, \bibinfo{person}{Bryan Russell}, \bibinfo{person}{Oliver Wang}, {and} \bibinfo{person}{Gautham~J. Mysore}.} \bibinfo{year}{2019}\natexlab{}.
\newblock \showarticletitle{B-Script: Transcript-based B-roll Video Editing with Recommendations}. In \bibinfo{booktitle}{\emph{Proceedings of the 2019 CHI Conference on Human Factors in Computing Systems}} (Glasgow, Scotland Uk) \emph{(\bibinfo{series}{CHI '19})}. \bibinfo{publisher}{Association for Computing Machinery}, \bibinfo{address}{New York, NY, USA}, \bibinfo{pages}{1–11}.
\newblock
\showISBNx{9781450359702}
\href{https://doi.org/10.1145/3290605.3300311}{doi:\nolinkurl{10.1145/3290605.3300311}}


\bibitem[Huh et~al\mbox{.}(2023)]%
        {10.1145/3544548.3581494}
\bibfield{author}{\bibinfo{person}{Mina Huh}, \bibinfo{person}{Saelyne Yang}, \bibinfo{person}{Yi-Hao Peng}, \bibinfo{person}{Xiang~'Anthony' Chen}, \bibinfo{person}{Young-Ho Kim}, {and} \bibinfo{person}{Amy Pavel}.} \bibinfo{year}{2023}\natexlab{}.
\newblock \showarticletitle{AVscript: Accessible Video Editing with Audio-Visual Scripts}. In \bibinfo{booktitle}{\emph{Proceedings of the 2023 CHI Conference on Human Factors in Computing Systems}} (Hamburg, Germany) \emph{(\bibinfo{series}{CHI '23})}. \bibinfo{publisher}{Association for Computing Machinery}, \bibinfo{address}{New York, NY, USA}, Article \bibinfo{articleno}{796}, \bibinfo{numpages}{17}~pages.
\newblock
\showISBNx{9781450394215}
\href{https://doi.org/10.1145/3544548.3581494}{doi:\nolinkurl{10.1145/3544548.3581494}}


\bibitem[Inc.({[n.\,d.]})]%
        {AdobePremierePro}
\bibfield{author}{\bibinfo{person}{Adobe Inc.}} \bibinfo{year}{[n.\,d.]}\natexlab{}.
\newblock \bibinfo{booktitle}{\emph{Adobe Premiere Pro: Video editing software}}.
\newblock Adobe Inc.
\newblock
\urldef\tempurl%
\url{https://www.adobe.com/products/premiere.html}
\showURL{%
\tempurl}


\bibitem[Kim et~al\mbox{.}(2024)]%
        {Hyuhng:2024:aligning}
\bibfield{author}{\bibinfo{person}{Hyuhng~Joon Kim}, \bibinfo{person}{Youna Kim}, \bibinfo{person}{Cheonbok Park}, \bibinfo{person}{Junyeob Kim}, \bibinfo{person}{Choonghyun Park}, \bibinfo{person}{Kang~Min Yoo}, \bibinfo{person}{Sang goo Lee}, {and} \bibinfo{person}{Taeuk Kim}.} \bibinfo{year}{2024}\natexlab{}.
\newblock \bibinfo{title}{Aligning Language Models to Explicitly Handle Ambiguity}.
\newblock
\showeprint[arxiv]{2404.11972}~[cs.CL]


\bibitem[Lee et~al\mbox{.}(2007)]%
        {Lee:2007:emotive}
\bibfield{author}{\bibinfo{person}{Daninel~G. Lee}, \bibinfo{person}{Deborah~I. Fels}, {and} \bibinfo{person}{John~Patrick Udo}.} \bibinfo{year}{2007}\natexlab{}.
\newblock \showarticletitle{Emotive captioning}.
\newblock \bibinfo{journal}{\emph{Computers in Entertainment}} \bibinfo{volume}{5}, \bibinfo{number}{2}, Article \bibinfo{articleno}{11} (\bibinfo{date}{April} \bibinfo{year}{2007}), \bibinfo{numpages}{15}~pages.
\newblock
\href{https://doi.org/10.1145/1279540.1279551}{doi:\nolinkurl{10.1145/1279540.1279551}}


\bibitem[Lee et~al\mbox{.}(2002)]%
        {Lee:2002:kinetic}
\bibfield{author}{\bibinfo{person}{Johnny~C. Lee}, \bibinfo{person}{Jodi Forlizzi}, {and} \bibinfo{person}{Scott~E. Hudson}.} \bibinfo{year}{2002}\natexlab{}.
\newblock \showarticletitle{The kinetic typography engine: an extensible system for animating expressive text}. In \bibinfo{booktitle}{\emph{Proceedings of the 15th Annual ACM Symposium on User Interface Software and Technology}} (Paris, France) \emph{(\bibinfo{series}{UIST '02})}. \bibinfo{publisher}{ACM}, \bibinfo{address}{New York, NY, USA}, \bibinfo{pages}{81–90}.
\newblock
\showISBNx{1581134886}
\href{https://doi.org/10.1145/571985.571997}{doi:\nolinkurl{10.1145/571985.571997}}


\bibitem[Li et~al\mbox{.}(2024)]%
        {Zelong:2024:formalllm}
\bibfield{author}{\bibinfo{person}{Zelong Li}, \bibinfo{person}{Wenyue Hua}, \bibinfo{person}{Hao Wang}, \bibinfo{person}{He Zhu}, {and} \bibinfo{person}{Yongfeng Zhang}.} \bibinfo{year}{2024}\natexlab{}.
\newblock \bibinfo{title}{Formal-LLM: Integrating Formal Language and Natural Language for Controllable LLM-based Agents}.
\newblock
\showeprint[arxiv]{2402.00798}~[cs.LG]


\bibitem[Liu et~al\mbox{.}(2024)]%
        {10.48550/arXiv.2408.13858}
\bibfield{author}{\bibinfo{person}{Minghao Liu}, \bibinfo{person}{Le Zhang}, \bibinfo{person}{Yingjie Tian}, \bibinfo{person}{Xiaochao Qu}, \bibinfo{person}{Luoqi Liu}, {and} \bibinfo{person}{Ting Liu}.} \bibinfo{year}{2024}\natexlab{}.
\newblock \bibinfo{title}{Draw Like an Artist: Complex Scene Generation with Diffusion Model via Composition, Painting, and Retouching}.
\newblock
\href{https://doi.org/10.48550/arXiv.2408.13858}{doi:\nolinkurl{10.48550/arXiv.2408.13858}}
\showeprint[arxiv]{2408.13858}~[cs.CV]


\bibitem[Liu et~al\mbox{.}(2023)]%
        {10.48550/arXiv.2303.04761}
\bibfield{author}{\bibinfo{person}{Shaoteng Liu}, \bibinfo{person}{Yuechen Zhang}, \bibinfo{person}{Wenbo Li}, \bibinfo{person}{Zhe Lin}, {and} \bibinfo{person}{Jiaya Jia}.} \bibinfo{year}{2023}\natexlab{}.
\newblock \bibinfo{title}{Video-P2P: Video Editing with Cross-attention Control}.
\newblock
\href{https://doi.org/10.48550/arXiv.2303.04761}{doi:\nolinkurl{10.48550/arXiv.2303.04761}}
\showeprint[arxiv]{2303.04761}~[cs.CV]


\bibitem[Ltd.({[n.\,d.]})]%
        {DaVinciResolve}
\bibfield{author}{\bibinfo{person}{Blackmagic Design~Pty. Ltd.}} \bibinfo{year}{[n.\,d.]}\natexlab{}.
\newblock \bibinfo{booktitle}{\emph{DaVinci Resolve: Professional Video Editing and Color Grading Software}}.
\newblock
\urldef\tempurl%
\url{https://www.blackmagicdesign.com/cn/products/davinciresolve}
\showURL{%
\tempurl}


\bibitem[MacNeil et~al\mbox{.}(2023)]%
        {Stephen:2023:prompt}
\bibfield{author}{\bibinfo{person}{Stephen MacNeil}, \bibinfo{person}{Andrew Tran}, \bibinfo{person}{Joanne Kim}, \bibinfo{person}{Ziheng Huang}, \bibinfo{person}{Seth Bernstein}, {and} \bibinfo{person}{Dan Mogil}.} \bibinfo{year}{2023}\natexlab{}.
\newblock \bibinfo{title}{Prompt Middleware: Mapping Prompts for Large Language Models to UI Affordances}.
\newblock
\showeprint[arxiv]{2307.01142}~[cs.HC]


\bibitem[Masson et~al\mbox{.}(2024)]%
        {10.1145/3613904.3642462}
\bibfield{author}{\bibinfo{person}{Damien Masson}, \bibinfo{person}{Sylvain Malacria}, \bibinfo{person}{G\'{e}ry Casiez}, {and} \bibinfo{person}{Daniel Vogel}.} \bibinfo{year}{2024}\natexlab{}.
\newblock \showarticletitle{DirectGPT: A Direct Manipulation Interface to Interact with Large Language Models}. In \bibinfo{booktitle}{\emph{Proceedings of the 2024 CHI Conference on Human Factors in Computing Systems}} (Honolulu, HI, USA) \emph{(\bibinfo{series}{CHI '24})}. \bibinfo{publisher}{Association for Computing Machinery}, \bibinfo{address}{New York, NY, USA}, Article \bibinfo{articleno}{975}, \bibinfo{numpages}{16}~pages.
\newblock
\showISBNx{9798400703300}
\href{https://doi.org/10.1145/3613904.3642462}{doi:\nolinkurl{10.1145/3613904.3642462}}


\bibitem[Murakami et~al\mbox{.}(2024)]%
        {10.1145/3613904.3642515}
\bibfield{author}{\bibinfo{person}{Taichi Murakami}, \bibinfo{person}{Kazuyuki Fujita}, \bibinfo{person}{Kotaro Hara}, \bibinfo{person}{Kazuki Takashima}, {and} \bibinfo{person}{Yoshifumi Kitamura}.} \bibinfo{year}{2024}\natexlab{}.
\newblock \showarticletitle{SwapVid: Integrating Video Viewing and Document Exploration with Direct Manipulation}. In \bibinfo{booktitle}{\emph{Proceedings of the 2024 CHI Conference on Human Factors in Computing Systems}} (Honolulu, HI, USA) \emph{(\bibinfo{series}{CHI '24})}. \bibinfo{publisher}{Association for Computing Machinery}, \bibinfo{address}{New York, NY, USA}, Article \bibinfo{articleno}{1035}, \bibinfo{numpages}{13}~pages.
\newblock
\showISBNx{9798400703300}
\href{https://doi.org/10.1145/3613904.3642515}{doi:\nolinkurl{10.1145/3613904.3642515}}


\bibitem[Myers(1999)]%
        {Myers:1999:cpef}
\bibfield{author}{\bibinfo{person}{Karen~L. Myers}.} \bibinfo{year}{1999}\natexlab{}.
\newblock \showarticletitle{CPEF: A Continuous Planning and Execution Framework}.
\newblock \bibinfo{journal}{\emph{AI Magazine}} \bibinfo{volume}{20}, \bibinfo{number}{4} (\bibinfo{date}{Dec.} \bibinfo{year}{1999}), \bibinfo{pages}{63}.
\newblock
\href{https://doi.org/10.1609/aimag.v20i4.1480}{doi:\nolinkurl{10.1609/aimag.v20i4.1480}}


\bibitem[Oh et~al\mbox{.}(2018)]%
        {10.1145/3173574.3174223}
\bibfield{author}{\bibinfo{person}{Changhoon Oh}, \bibinfo{person}{Jungwoo Song}, \bibinfo{person}{Jinhan Choi}, \bibinfo{person}{Seonghyeon Kim}, \bibinfo{person}{Sungwoo Lee}, {and} \bibinfo{person}{Bongwon Suh}.} \bibinfo{year}{2018}\natexlab{}.
\newblock \showarticletitle{I Lead, You Help but Only with Enough Details: Understanding User Experience of Co-Creation with Artificial Intelligence}. In \bibinfo{booktitle}{\emph{Proceedings of the 2018 CHI Conference on Human Factors in Computing Systems}} (Montreal QC, Canada) \emph{(\bibinfo{series}{CHI '18})}. \bibinfo{publisher}{Association for Computing Machinery}, \bibinfo{address}{New York, NY, USA}, \bibinfo{pages}{1–13}.
\newblock
\showISBNx{9781450356206}
\href{https://doi.org/10.1145/3173574.3174223}{doi:\nolinkurl{10.1145/3173574.3174223}}


\bibitem[Park et~al\mbox{.}(2024)]%
        {Park:2024:clara}
\bibfield{author}{\bibinfo{person}{Jeongeun Park}, \bibinfo{person}{Seungwon Lim}, \bibinfo{person}{Joonhyung Lee}, \bibinfo{person}{Sangbeom Park}, \bibinfo{person}{Minsuk Chang}, \bibinfo{person}{Youngjae Yu}, {and} \bibinfo{person}{Sungjoon Choi}.} \bibinfo{year}{2024}\natexlab{}.
\newblock \showarticletitle{CLARA: Classifying and Disambiguating User Commands for Reliable Interactive Robotic Agents}.
\newblock \bibinfo{journal}{\emph{IEEE Robotics and Automation Letters}} \bibinfo{volume}{9}, \bibinfo{number}{2} (\bibinfo{year}{2024}), \bibinfo{pages}{1059--1066}.
\newblock
\href{https://doi.org/10.1109/LRA.2023.3338514}{doi:\nolinkurl{10.1109/LRA.2023.3338514}}


\bibitem[Pavel et~al\mbox{.}(2020)]%
        {10.1145/3379337.3415864}
\bibfield{author}{\bibinfo{person}{Amy Pavel}, \bibinfo{person}{Gabriel Reyes}, {and} \bibinfo{person}{Jeffrey~P. Bigham}.} \bibinfo{year}{2020}\natexlab{}.
\newblock \showarticletitle{Rescribe: Authoring and Automatically Editing Audio Descriptions}. In \bibinfo{booktitle}{\emph{Proceedings of the 33rd Annual ACM Symposium on User Interface Software and Technology}} (Virtual Event, USA) \emph{(\bibinfo{series}{UIST '20})}. \bibinfo{publisher}{Association for Computing Machinery}, \bibinfo{address}{New York, NY, USA}, \bibinfo{pages}{747–759}.
\newblock
\showISBNx{9781450375146}
\href{https://doi.org/10.1145/3379337.3415864}{doi:\nolinkurl{10.1145/3379337.3415864}}


\bibitem[Raees et~al\mbox{.}(2024)]%
        {10.1016/j.ijhcs.2024.103301}
\bibfield{author}{\bibinfo{person}{Muhammad Raees}, \bibinfo{person}{Inge Meijerink}, \bibinfo{person}{Ioanna Lykourentzou}, \bibinfo{person}{Vassilis-Javed Khan}, {and} \bibinfo{person}{Konstantinos Papangelis}.} \bibinfo{year}{2024}\natexlab{}.
\newblock \showarticletitle{From explainable to interactive AI: A literature review on current trends in human-AI interaction}.
\newblock \bibinfo{journal}{\emph{International Journal of Human-Computer Studies}}  \bibinfo{volume}{189} (\bibinfo{year}{2024}), \bibinfo{pages}{103301}.
\newblock
\showISSN{1071-5819}
\href{https://doi.org/10.1016/j.ijhcs.2024.103301}{doi:\nolinkurl{10.1016/j.ijhcs.2024.103301}}


\bibitem[Rao et~al\mbox{.}(2024)]%
        {10.48550/arXiv.2410.03224}
\bibfield{author}{\bibinfo{person}{Anyi Rao}, \bibinfo{person}{Jean-Peïc Chou}, {and} \bibinfo{person}{Maneesh Agrawala}.} \bibinfo{year}{2024}\natexlab{}.
\newblock \bibinfo{title}{ScriptViz: A Visualization Tool to Aid Scriptwriting based on a Large Movie Database}.
\newblock
\href{https://doi.org/10.48550/arXiv.2410.03224}{doi:\nolinkurl{10.48550/arXiv.2410.03224}}
\showeprint[arxiv]{2410.03224}~[cs.HC]


\bibitem[{Runway}({[n.\,d.]})]%
        {RunwayFrames}
\bibfield{author}{\bibinfo{person}{{Runway}}.} \bibinfo{year}{[n.\,d.]}\natexlab{}.
\newblock \bibinfo{booktitle}{\emph{Runway: Tools for Human Imagination}}.
\newblock
\urldef\tempurl%
\url{https://runwayml.com/}
\showURL{%
\tempurl}


\bibitem[Shao et~al\mbox{.}(2025)]%
        {Shao:2025:animation}
\bibfield{author}{\bibinfo{person}{Zekai Shao}, \bibinfo{person}{Leixian Shen}, \bibinfo{person}{Haotian Li}, \bibinfo{person}{Yi Shan}, \bibinfo{person}{Huamin Qu}, \bibinfo{person}{Yun Wang}, {and} \bibinfo{person}{Siming Chen}.} \bibinfo{year}{2025}\natexlab{}.
\newblock \showarticletitle{Narrative Player: Reviving Data Narratives with Visuals}.
\newblock \bibinfo{journal}{\emph{IEEE Transactions on Visualization and Computer Graphics}} (\bibinfo{year}{2025}), \bibinfo{pages}{1--15}.
\newblock
\href{https://doi.org/10.1109/TVCG.2025.3530512}{doi:\nolinkurl{10.1109/TVCG.2025.3530512}}


\bibitem[Shen et~al\mbox{.}(2024a)]%
        {10.1109/GEN4DS63889.2024.00008}
\bibfield{author}{\bibinfo{person}{Leixian Shen}, \bibinfo{person}{Haotian Li}, \bibinfo{person}{Yun Wang}, {and} \bibinfo{person}{Huamin Qu}.} \bibinfo{year}{2024}\natexlab{a}.
\newblock \showarticletitle{From Data to Story: Towards Automatic Animated Data Video Creation with LLM-Based Multi-Agent Systems}. In \bibinfo{booktitle}{\emph{2024 IEEE VIS Workshop on Data Storytelling in an Era of Generative AI (GEN4DS)}}. \bibinfo{pages}{20--27}.
\newblock
\href{https://doi.org/10.1109/GEN4DS63889.2024.00008}{doi:\nolinkurl{10.1109/GEN4DS63889.2024.00008}}


\bibitem[Shen et~al\mbox{.}(2024b)]%
        {10.1109/TVCG.2023.3327197}
\bibfield{author}{\bibinfo{person}{Leixian Shen}, \bibinfo{person}{Yizhi Zhang}, \bibinfo{person}{Haidong Zhang}, {and} \bibinfo{person}{Yun Wang}.} \bibinfo{year}{2024}\natexlab{b}.
\newblock \showarticletitle{Data Player: Automatic Generation of Data Videos with Narration-Animation Interplay}.
\newblock \bibinfo{journal}{\emph{IEEE Transactions on Visualization and Computer Graphics}} \bibinfo{volume}{30}, \bibinfo{number}{1} (\bibinfo{year}{2024}), \bibinfo{pages}{109--119}.
\newblock
\href{https://doi.org/10.1109/TVCG.2023.3327197}{doi:\nolinkurl{10.1109/TVCG.2023.3327197}}


\bibitem[Shneiderman(1981)]%
        {10.1145/1015579.810991}
\bibfield{author}{\bibinfo{person}{Ben Shneiderman}.} \bibinfo{year}{1981}\natexlab{}.
\newblock \showarticletitle{Direct manipulation: A step beyond programming languages (abstract only)}.
\newblock \bibinfo{journal}{\emph{SIGSOC Bull.}} \bibinfo{volume}{13}, \bibinfo{number}{2–3} (\bibinfo{date}{May} \bibinfo{year}{1981}), \bibinfo{pages}{143}.
\newblock
\showISSN{0163-5794}
\href{https://doi.org/10.1145/1015579.810991}{doi:\nolinkurl{10.1145/1015579.810991}}


\bibitem[Shneiderman(1982)]%
        {10.1080/01449298208914450}
\bibfield{author}{\bibinfo{person}{Ben Shneiderman}.} \bibinfo{year}{1982}\natexlab{}.
\newblock \showarticletitle{The future of interactive systems and the emergence of direct manipulation}.
\newblock \bibinfo{journal}{\emph{Behaviour \& Information Technology}} \bibinfo{volume}{1}, \bibinfo{number}{3} (\bibinfo{year}{1982}), \bibinfo{pages}{237--256}.
\newblock
\href{https://doi.org/10.1080/01449298208914450}{doi:\nolinkurl{10.1080/01449298208914450}}


\bibitem[Siddiqui et~al\mbox{.}(2025)]%
        {10.48550/arXiv.2502.10638}
\bibfield{author}{\bibinfo{person}{Momin Siddiqui}, \bibinfo{person}{Roy Pea}, {and} \bibinfo{person}{Hari Subramonyam}.} \bibinfo{year}{2025}\natexlab{}.
\newblock \bibinfo{title}{Script\&Shift: A Layered Interface Paradigm for Integrating Content Development and Rhetorical Strategy with LLM Writing Assistants}.
\newblock
\href{https://doi.org/10.48550/arXiv.2502.10638}{doi:\nolinkurl{10.48550/arXiv.2502.10638}}
\showeprint[arxiv]{2502.10638}~[cs.HC]


\bibitem[Sowa et~al\mbox{.}(2021)]%
        {10.1016/j.jbusres.2020.11.038}
\bibfield{author}{\bibinfo{person}{Konrad Sowa}, \bibinfo{person}{Aleksandra Przegalinska}, {and} \bibinfo{person}{Leon Ciechanowski}.} \bibinfo{year}{2021}\natexlab{}.
\newblock \showarticletitle{Cobots in knowledge work: Human – AI collaboration in managerial professions}.
\newblock \bibinfo{journal}{\emph{Journal of Business Research}}  \bibinfo{volume}{125} (\bibinfo{year}{2021}), \bibinfo{pages}{135--142}.
\newblock
\showISSN{0148-2963}
\href{https://doi.org/10.1016/j.jbusres.2020.11.038}{doi:\nolinkurl{10.1016/j.jbusres.2020.11.038}}


\bibitem[Strobelt et~al\mbox{.}(2023)]%
        {Strobelt:2023:interactive}
\bibfield{author}{\bibinfo{person}{Hendrik Strobelt}, \bibinfo{person}{Albert Webson}, \bibinfo{person}{Victor Sanh}, \bibinfo{person}{Benjamin Hoover}, \bibinfo{person}{Johanna Beyer}, \bibinfo{person}{Hanspeter Pfister}, {and} \bibinfo{person}{Alexander~M. Rush}.} \bibinfo{year}{2023}\natexlab{}.
\newblock \showarticletitle{Interactive and Visual Prompt Engineering for Ad-hoc Task Adaptation with Large Language Models}.
\newblock \bibinfo{journal}{\emph{IEEE Transactions on Visualization and Computer Graphics}} \bibinfo{volume}{29}, \bibinfo{number}{1} (\bibinfo{year}{2023}), \bibinfo{pages}{1146--1156}.
\newblock
\href{https://doi.org/10.1109/TVCG.2022.3209479}{doi:\nolinkurl{10.1109/TVCG.2022.3209479}}


\bibitem[Sun et~al\mbox{.}(2024)]%
        {10.48550/arXiv.2407.07111}
\bibfield{author}{\bibinfo{person}{Wenhao Sun}, \bibinfo{person}{Rong-Cheng Tu}, \bibinfo{person}{Jingyi Liao}, {and} \bibinfo{person}{Dacheng Tao}.} \bibinfo{year}{2024}\natexlab{}.
\newblock \bibinfo{title}{Diffusion Model-Based Video Editing: A Survey}.
\newblock
\href{https://doi.org/10.48550/arXiv.2407.07111}{doi:\nolinkurl{10.48550/arXiv.2407.07111}}
\showeprint[arxiv]{2407.07111}~[cs.CV]


\bibitem[Torricelli et~al\mbox{.}(2024)]%
        {Torricelli:2024:theRole}
\bibfield{author}{\bibinfo{person}{Maddalena Torricelli}, \bibinfo{person}{Mauro Martino}, \bibinfo{person}{Andrea Baronchelli}, {and} \bibinfo{person}{Luca~Maria Aiello}.} \bibinfo{year}{2024}\natexlab{}.
\newblock \showarticletitle{The Role of Interface Design on Prompt-mediated Creativity in Generative AI}. In \bibinfo{booktitle}{\emph{Proceedings of the 16th ACM Web Science Conference}} (Stuttgart, Germany) \emph{(\bibinfo{series}{WEBSCI '24})}. \bibinfo{publisher}{Association for Computing Machinery}, \bibinfo{address}{New York, NY, USA}, \bibinfo{pages}{235–240}.
\newblock
\showISBNx{9798400703348}
\href{https://doi.org/10.1145/3614419.3644000}{doi:\nolinkurl{10.1145/3614419.3644000}}


\bibitem[Vijayvargiya et~al\mbox{.}(2025)]%
        {Sanidhya:2025:interactive}
\bibfield{author}{\bibinfo{person}{Sanidhya Vijayvargiya}, \bibinfo{person}{Xuhui Zhou}, \bibinfo{person}{Akhila Yerukola}, \bibinfo{person}{Maarten Sap}, {and} \bibinfo{person}{Graham Neubig}.} \bibinfo{year}{2025}\natexlab{}.
\newblock \bibinfo{title}{Interactive Agents to Overcome Ambiguity in Software Engineering}.
\newblock
\showeprint[arxiv]{2502.13069}~[cs.AI]


\bibitem[V\"{o}ssing et~al\mbox{.}(2022)]%
        {10.1007/s10796-022-10284-3}
\bibfield{author}{\bibinfo{person}{Michael V\"{o}ssing}, \bibinfo{person}{Niklas K\"{u}hl}, \bibinfo{person}{Matteo Lind}, {and} \bibinfo{person}{Gerhard Satzger}.} \bibinfo{year}{2022}\natexlab{}.
\newblock \showarticletitle{Designing Transparency for Effective Human-AI Collaboration}.
\newblock \bibinfo{journal}{\emph{Information Systems Frontiers}} \bibinfo{volume}{24}, \bibinfo{number}{3} (\bibinfo{date}{June} \bibinfo{year}{2022}), \bibinfo{pages}{877–895}.
\newblock
\showISSN{1387-3326}
\href{https://doi.org/10.1007/s10796-022-10284-3}{doi:\nolinkurl{10.1007/s10796-022-10284-3}}


\bibitem[Wang et~al\mbox{.}(2024a)]%
        {10.1145/3640543.3645143}
\bibfield{author}{\bibinfo{person}{Bryan Wang}, \bibinfo{person}{Yuliang Li}, \bibinfo{person}{Zhaoyang Lv}, \bibinfo{person}{Haijun Xia}, \bibinfo{person}{Yan Xu}, {and} \bibinfo{person}{Raj Sodhi}.} \bibinfo{year}{2024}\natexlab{a}.
\newblock \showarticletitle{LAVE: LLM-Powered Agent Assistance and Language Augmentation for Video Editing}. In \bibinfo{booktitle}{\emph{Proceedings of the 29th International Conference on Intelligent User Interfaces}} (Greenville, SC, USA) \emph{(\bibinfo{series}{IUI '24})}. \bibinfo{publisher}{Association for Computing Machinery}, \bibinfo{address}{New York, NY, USA}, \bibinfo{pages}{699–714}.
\newblock
\showISBNx{9798400705083}
\href{https://doi.org/10.1145/3640543.3645143}{doi:\nolinkurl{10.1145/3640543.3645143}}


\bibitem[Wang et~al\mbox{.}(2024b)]%
        {10.1145/3613904.3642868}
\bibfield{author}{\bibinfo{person}{Sitong Wang}, \bibinfo{person}{Samia Menon}, \bibinfo{person}{Tao Long}, \bibinfo{person}{Keren Henderson}, \bibinfo{person}{Dingzeyu Li}, \bibinfo{person}{Kevin Crowston}, \bibinfo{person}{Mark Hansen}, \bibinfo{person}{Jeffrey~V Nickerson}, {and} \bibinfo{person}{Lydia~B Chilton}.} \bibinfo{year}{2024}\natexlab{b}.
\newblock \showarticletitle{ReelFramer: Human-AI Co-Creation for News-to-Video Translation}. In \bibinfo{booktitle}{\emph{Proceedings of the 2024 CHI Conference on Human Factors in Computing Systems}} (Honolulu, HI, USA) \emph{(\bibinfo{series}{CHI '24})}. \bibinfo{publisher}{Association for Computing Machinery}, \bibinfo{address}{New York, NY, USA}, Article \bibinfo{articleno}{169}, \bibinfo{numpages}{20}~pages.
\newblock
\showISBNx{9798400703300}
\href{https://doi.org/10.1145/3613904.3642868}{doi:\nolinkurl{10.1145/3613904.3642868}}


\bibitem[Wang et~al\mbox{.}(2024c)]%
        {10.1145/3643834.3661591}
\bibfield{author}{\bibinfo{person}{Sitong Wang}, \bibinfo{person}{Zheng Ning}, \bibinfo{person}{Anh Truong}, \bibinfo{person}{Mira Dontcheva}, \bibinfo{person}{Dingzeyu Li}, {and} \bibinfo{person}{Lydia~B Chilton}.} \bibinfo{year}{2024}\natexlab{c}.
\newblock \showarticletitle{PodReels: Human-AI Co-Creation of Video Podcast Teasers}. In \bibinfo{booktitle}{\emph{Proceedings of the 2024 ACM Designing Interactive Systems Conference}} (Copenhagen, Denmark) \emph{(\bibinfo{series}{DIS '24})}. \bibinfo{publisher}{Association for Computing Machinery}, \bibinfo{address}{New York, NY, USA}, \bibinfo{pages}{958–974}.
\newblock
\showISBNx{9798400705830}
\href{https://doi.org/10.1145/3643834.3661591}{doi:\nolinkurl{10.1145/3643834.3661591}}


\bibitem[Wang et~al\mbox{.}(2024d)]%
        {10.1109/TVCG.2024.3411575}
\bibfield{author}{\bibinfo{person}{Yun Wang}, \bibinfo{person}{Leixian Shen}, \bibinfo{person}{Zhengxin You}, \bibinfo{person}{Xinhuan Shu}, \bibinfo{person}{Bongshin Lee}, \bibinfo{person}{John Thompson}, \bibinfo{person}{Haidong Zhang}, {and} \bibinfo{person}{Dongmei Zhang}.} \bibinfo{year}{2024}\natexlab{d}.
\newblock \showarticletitle{WonderFlow: Narration-Centric Design of Animated Data Videos}.
\newblock \bibinfo{journal}{\emph{IEEE Transactions on Visualization and Computer Graphics}} (\bibinfo{year}{2024}), \bibinfo{pages}{1--17}.
\newblock
\href{https://doi.org/10.1109/TVCG.2024.3411575}{doi:\nolinkurl{10.1109/TVCG.2024.3411575}}


\bibitem[Wu et~al\mbox{.}(2021)]%
        {10.1007/978-3-030-78462-1_13}
\bibfield{author}{\bibinfo{person}{Zhuohao Wu}, \bibinfo{person}{Danwen Ji}, \bibinfo{person}{Kaiwen Yu}, \bibinfo{person}{Xianxu Zeng}, \bibinfo{person}{Dingming Wu}, {and} \bibinfo{person}{Mohammad Shidujaman}.} \bibinfo{year}{2021}\natexlab{}.
\newblock \showarticletitle{AI Creativity and the Human-AI Co-creation Model}. In \bibinfo{booktitle}{\emph{Human-Computer Interaction. Theory, Methods and Tools: Thematic Area, HCI 2021, Held as Part of the 23rd HCI International Conference, HCII 2021, Virtual Event, July 24–29, 2021, Proceedings, Part I}}. \bibinfo{publisher}{Springer-Verlag}, \bibinfo{address}{Berlin, Heidelberg}, \bibinfo{pages}{171–190}.
\newblock
\showISBNx{978-3-030-78461-4}
\href{https://doi.org/10.1007/978-3-030-78462-1_13}{doi:\nolinkurl{10.1007/978-3-030-78462-1_13}}


\bibitem[Xie et~al\mbox{.}(2024)]%
        {Xie:2024:wordcloud}
\bibfield{author}{\bibinfo{person}{Liwenhan Xie}, \bibinfo{person}{Xinhuan Shu}, \bibinfo{person}{Jeon~Cheol Su}, \bibinfo{person}{Yun Wang}, \bibinfo{person}{Siming Chen}, {and} \bibinfo{person}{Huamin Qu}.} \bibinfo{year}{2024}\natexlab{}.
\newblock \showarticletitle{Creating Emordle: Animating Word Cloud for Emotion Expression}.
\newblock \bibinfo{journal}{\emph{IEEE Transactions on Visualization and Computer Graphics}} \bibinfo{volume}{30}, \bibinfo{number}{8} (\bibinfo{year}{2024}), \bibinfo{pages}{5198--5211}.
\newblock
\href{https://doi.org/10.1109/TVCG.2023.3286392}{doi:\nolinkurl{10.1109/TVCG.2023.3286392}}


\bibitem[Xie et~al\mbox{.}(2023)]%
        {Xie:2023:wakey}
\bibfield{author}{\bibinfo{person}{Liwenhan Xie}, \bibinfo{person}{Zhaoyu Zhou}, \bibinfo{person}{Kerun Yu}, \bibinfo{person}{Yun Wang}, \bibinfo{person}{Huamin Qu}, {and} \bibinfo{person}{Siming Chen}.} \bibinfo{year}{2023}\natexlab{}.
\newblock \showarticletitle{Wakey-Wakey: Animate Text by Mimicking Characters in a GIF}. In \bibinfo{booktitle}{\emph{Proceedings of the 36th Annual ACM Symposium on User Interface Software and Technology}} (San Francisco, CA, USA) \emph{(\bibinfo{series}{UIST '23})}. \bibinfo{publisher}{ACM}, \bibinfo{address}{New York, NY, USA}, Article \bibinfo{articleno}{98}, \bibinfo{numpages}{14}~pages.
\newblock
\showISBNx{9798400701320}
\href{https://doi.org/10.1145/3586183.3606813}{doi:\nolinkurl{10.1145/3586183.3606813}}


\bibitem[Xing et~al\mbox{.}(2024)]%
        {10.1145/3696415}
\bibfield{author}{\bibinfo{person}{Zhen Xing}, \bibinfo{person}{Qijun Feng}, \bibinfo{person}{Haoran Chen}, \bibinfo{person}{Qi Dai}, \bibinfo{person}{Han Hu}, \bibinfo{person}{Hang Xu}, \bibinfo{person}{Zuxuan Wu}, {and} \bibinfo{person}{Yu-Gang Jiang}.} \bibinfo{year}{2024}\natexlab{}.
\newblock \showarticletitle{A Survey on Video Diffusion Models}.
\newblock \bibinfo{journal}{\emph{ACM Comput. Surv.}} \bibinfo{volume}{57}, \bibinfo{number}{2}, Article \bibinfo{articleno}{41} (\bibinfo{date}{Nov.} \bibinfo{year}{2024}), \bibinfo{numpages}{42}~pages.
\newblock
\showISSN{0360-0300}
\href{https://doi.org/10.1145/3696415}{doi:\nolinkurl{10.1145/3696415}}


\bibitem[Yang et~al\mbox{.}(2023)]%
        {10.1145/3626235}
\bibfield{author}{\bibinfo{person}{Ling Yang}, \bibinfo{person}{Zhilong Zhang}, \bibinfo{person}{Yang Song}, \bibinfo{person}{Shenda Hong}, \bibinfo{person}{Runsheng Xu}, \bibinfo{person}{Yue Zhao}, \bibinfo{person}{Wentao Zhang}, \bibinfo{person}{Bin Cui}, {and} \bibinfo{person}{Ming-Hsuan Yang}.} \bibinfo{year}{2023}\natexlab{}.
\newblock \showarticletitle{Diffusion Models: A Comprehensive Survey of Methods and Applications}.
\newblock \bibinfo{journal}{\emph{ACM Comput. Surv.}} \bibinfo{volume}{56}, \bibinfo{number}{4}, Article \bibinfo{articleno}{105} (\bibinfo{date}{Nov.} \bibinfo{year}{2023}), \bibinfo{numpages}{39}~pages.
\newblock
\showISSN{0360-0300}
\href{https://doi.org/10.1145/3626235}{doi:\nolinkurl{10.1145/3626235}}


\bibitem[Yeo(2008)]%
        {Yeo:2008:emotional}
\bibfield{author}{\bibinfo{person}{Zhiquan Yeo}.} \bibinfo{year}{2008}\natexlab{}.
\newblock \showarticletitle{Emotional instant messaging with KIM}. In \bibinfo{booktitle}{\emph{CHI '08 Extended Abstracts on Human Factors in Computing Systems}} (Florence, Italy) \emph{(\bibinfo{series}{CHI EA '08})}. \bibinfo{publisher}{ACM}, \bibinfo{address}{New York, NY, USA}, \bibinfo{pages}{3729–3734}.
\newblock
\showISBNx{9781605580128}
\href{https://doi.org/10.1145/1358628.1358921}{doi:\nolinkurl{10.1145/1358628.1358921}}


\bibitem[Zhao et~al\mbox{.}(2024)]%
        {Zhao:2024:lightva}
\bibfield{author}{\bibinfo{person}{Yuheng Zhao}, \bibinfo{person}{Junjie Wang}, \bibinfo{person}{Linbin Xiang}, \bibinfo{person}{Xiaowen Zhang}, \bibinfo{person}{Zifei Guo}, \bibinfo{person}{Cagatay Turkay}, \bibinfo{person}{Yu Zhang}, {and} \bibinfo{person}{Siming Chen}.} \bibinfo{year}{2024}\natexlab{}.
\newblock \showarticletitle{LightVA: Lightweight Visual Analytics with LLM Agent-Based Task Planning and Execution}.
\newblock \bibinfo{journal}{\emph{IEEE Transactions on Visualization and Computer Graphics}} (\bibinfo{year}{2024}), \bibinfo{pages}{1--13}.
\newblock
\href{https://doi.org/10.1109/TVCG.2024.3496112}{doi:\nolinkurl{10.1109/TVCG.2024.3496112}}


\end{thebibliography}

\end{document}